\documentclass[11pt]{article}
\pdfoutput=1
\usepackage{amsmath,amsfonts,feynmp,epsf}
\usepackage{amssymb}
\usepackage{graphicx}
\usepackage{grffile}
\input epsf
\usepackage[nosort]{cite}

\usepackage{epstopdf}
\usepackage{enumerate}
\usepackage{tensor}
\usepackage{mathrsfs}

\usepackage[english]{babel}
\usepackage{braket}
\usepackage{feynman}
\usepackage{dcolumn}
\usepackage{bm}
\usepackage{hyperref}
\usepackage{slashed}

\usepackage{comment}

\makeatletter\@addtoreset{equation}{section}\makeatother

\def\ket#1{\mathinner{|{#1}\rangle}}

{\catcode`\|=\active
  \gdef\Braket#1{\left<\mathcode`\|"8000\let|\BraVert {#1}\right>}}
\def\BraVert{\egroup\,\mid@vertical\,\bgroup}
{\catcode`\|=\active
  \gdef\set#1{\mathinner{\lbrace\,{\mathcode`\|"8000\let|\midvert #1}\,\rbrace}}
  \gdef\Set#1{\left\{\:{\mathcode`\|"8000\let|\SetVert #1}\:\right\}}}
\def\midvert{\egroup\mid\bgroup}
\def\SetVert{\egroup\;\mid@vertical\;\bgroup}

\def\be{\begin{equation} }
\def\ee{ \end{equation} }
\def\bea{\begin{eqnarray}}
\def\eea{\end{eqnarray}}
\def\ie{\begin{equation}\begin{aligned}}
\def\fe{\end{aligned}\end{equation}}

\makeatletter\@addtoreset{equation}{section}\makeatother

\newcommand{\preprint}[1]{\begin{table}[t]  
             \begin{flushright}               
             {#1}                             
             \end{flushright}                 
             \end{table}}                     
\renewcommand{\title}[1]{\vbox{\center\LARGE{#1}}\vspace{5mm}}
\renewcommand{\author}[1]{\vbox{\center#1}\vspace{5mm}}
\newcommand{\address}[1]{\vbox{\center\em#1}}

\newcommand{\deh}{\hat{\partial}}

\newcommand{\Tr}{\text{tr~}}
\newcommand{\tr}{\text{tr~}}

\begin{document}
\begin{titlepage}
\preprint{PUPT-2512}
\preprint{LMU-ASC 52/16}
\begin{center}
\vskip 1cm

\title{On the Higher-Spin Spectrum \\ in Large $N$ Chern-Simons Vector Models}

\author{S. Giombi$^{1}$, V. Gurucharan$^{2}$, V. Kirilin$^{1,5}$, S. Prakash$^{2}$, E. Skvortsov$^{3,4}$}

\address{${}^1$Department of Physics, Princeton University, Princeton, NJ 08544}
\address{${}^2$Department of Physics and Computer Science, Dayalbagh Educational Institute, Dayalbagh, Agra, India 282005}
\address{${}^3$Arnold Sommerfeld Center for Theoretical Physics, Ludwig-Maximilians University Munich, Theresienstr. 37, D-80333 Munich, Germany}
\address{${}^4$Lebedev Institute of Physics, Leninsky ave. 53, 119991 Moscow, Russia}
\address{${}^5$ITEP, B. Cheremushkinskaya 25, Moscow, 117218, Russia}


\end{center}

\abstract{Chern-Simons gauge theories coupled to massless fundamental scalars or fermions define interesting non-supersymmetric 3d CFTs that 
possess approximate higher-spin symmetries at large $N$. In this paper, we compute the scaling 
dimensions of the higher-spin operators in these models, to leading order in the $1/N$ expansion and exactly in 
the 't~Hooft coupling. We obtain these results in two independent ways: by using conformal symmetry and the classical 
equations of motion to fix the structure of the current non-conservation, and by a direct Feynman diagram calculation. 
The full dependence on the 't~Hooft coupling can be restored by using results that follow from the weakly broken higher-spin symmetry. This analysis also allows us to obtain some explicit results for the non-conserved, parity-breaking structures that appear
in planar three-point functions of the higher-spin operators. At large 
spin, we find that the anomalous dimensions grow logarithmically with the spin, in agreement with general expectations. This logarithmic 
behavior disappears in the strong coupling limit, where the anomalous dimensions turn into those of the critical $O(N)$ or Gross-Neveu models, 
in agreement with the conjectured 3d bosonization duality. }

\vfill

\end{titlepage}

\eject \tableofcontents

\section{\label{sec:level1}Introduction and Summary}
Chern-Simons (CS) gauge theories coupled to massless matter fields lead to a large class of conformal field theories in three dimensions, with or without 
supersymmetry. A particularly interesting non-supersymmetric example is obtained by coupling a $U(N)$ (or $O(N)$) CS gauge theory to a fermion or scalar in 
the fundamental representation \cite{Giombi:2011kc, Aharony:2011jz}. The Chern-Simons coupling $k$ is quantized and cannot run (up to a possible integer shift at one loop). Therefore, in the fermionic case it is sufficient to tune away the relevant mass 
term to obtain a conformal field theory (CFT) for any $N$ and $k$ \cite{Giombi:2011kc}.\footnote{The level $k$ has to be half-integer due to the parity anomaly \cite{Niemi:1983rq, Redlich:1983dv, Redlich:1983kn}.} In the 
scalar case, one has a classically marginal coupling $\phi^6$ that can get generated along RG flow, but in the presence of CS interactions one can find zeroes of its beta function, at least for sufficiently large $N$ \cite{Aharony:2011jz}. One may also obtain ``critical" versions of these models by adding quartic self-interactions for the fundamental matter fields. In the scalar case, this leads to an IR fixed point which is a generalization of the familiar critical $O(N)$ model. In the fermionic case, at least in the large $N$ expansion, one finds UV fixed points which generalize the critical 3d Gross-Neveu model. 

The CFTs described above may be viewed as generalizations of the well-known bosonic and fermionic vector models by the addition of CS interactions, and we may refer to them 
as ``Chern-Simons vector models". Their investigation was initially motivated by the study of the AdS/CFT duality between Vasiliev higher-spin theory in AdS$_4$ 
\cite{Vasiliev:1992av}\footnote{See for instance \cite{Vasiliev:1999ba,Didenko:2014dwa,Giombi:2016ejx} for a review of the $4d$ Vasiliev equations.} and 
free/critical vectorial CFTs with scalar or fermionic fields \cite{Klebanov:2002ja,Leigh:2003gk,Sezgin:2003pt}. Gauging 
the global symmetries of the vector model by means of the CS gauge theory leads to a natural way to implement the singlet constraint, 
which is necessary in the conjecture of \cite{Klebanov:2002ja}. Remarkably, it turns out that in the 
't~Hooft limit of large $N$ with $\lambda=N/k$ fixed, the CS vector models admit an approximate higher-spin (HS) symmetry, similarly to their ungauged versions, in the sense that the currents $j_s$ are approximately conserved and have small anomalous dimension at large $N$ \cite{Giombi:2011kc, Aharony:2011jz}.  The fact that the anomalous dimensions are generated through $1/N$ corrections implies that the holographic dual to the CS vector models should be a parity breaking version of Vasiliev HS gravity, where the HS fields are classically massless, and masses are generated via bulk loop diagrams. The bulk HS theory is characterized by a parity breaking phase $\theta_0$, which is mapped to the CFT 't~Hooft coupling $\lambda$. See e.g. \cite{Chang:2012kt, Giombi:2012ms, Giombi:2016ejx} for reviews of this duality. 

A variety of new techniques have been developed and applied recently to the study of bosonic and fermionic vector models
\cite{Rychkov:2015naa,Basu:2015gpa,Sen:2015doa,Ghosh:2015opa,Raju:2015fza,Manashov:2015fha,Skvortsov:2015pea,Giombi:2016hkj,Hikida:2016wqj,Dey:2016zbg,
Bashmakov:2016uqk, Bashmakov:2016pcg, Nii:2016lpa,Gopakumar:2016wkt,Hikida:2016cla, Manashov:2016uam}, and bootstrap methods have also been applied for studying operators with large spin, e.g. \cite{Fitzpatrick:2012yx, Komargodski:2012ek, Kaviraj:2015cxa, Alday:2015eya, Kaviraj:2015xsa,Alday:2015ota}. Partially motivated by this body of works, we study the spectrum of $1/N$ scaling dimensions of single-trace, primary operators with $s\geq 1$ in Chern-Simons vector models.

As we review in section \ref{CS-actions}, the spectrum of single-trace primary operators in these models is very simple: it just consists of bilinears in the fundamental matter fields. These include a scalar operator ($\bar\phi\phi$ or $\bar\psi\psi$), and a tower of spinning operators $j_s$ of all integer spins. Owing to the topological nature of the CS gauge field, the addition of the CS interactions does not lead to any new local operator on top of the bilinears. It follows, as will be reviewed in more detail below, that the non-conservation of the HS currents $j_s$ must take the schematic form \cite{Giombi:2011kc, Aharony:2011jz, MZ}
\begin{equation}
\label{non-cons-intro}
\partial \cdot j_s \sim \sum_{s_1,s_2} \frac{1}{N} f^{(3)}_{s,s_1,s_2}(\lambda) \partial^n j_{s_1} \partial^m j_{s_2} +  
\sum_{s_1,s_2,s_3} \frac{1}{N^2} f^{(4)}_{s,s_1,s_2,s_3}(\lambda) \partial^n j_{s_1} \partial^m j_{s_2}  \partial^p j_{s_3} \,, 
\end{equation}
where the ``double-trace" and ``triple-trace" operators on the right-hand side correspond to products of the bilinears and their derivatives, and no ``single-trace" operator can appear, since there are none in the spectrum with the correct quantum numbers. The weakly broken HS symmetries corresponding to (\ref{non-cons-intro}) can be used to constrain all planar 2-point and 3-point functions of the single-trace operators in terms of two parameters \cite{MZ}.\footnote{In the case of regular CS-scalar theory or critical 
CS-fermion theory, there is an additional marginal parameter corresponding to sextic couplings.} The non-conservation equation (\ref{non-cons-intro}) 
also encodes the anomalous dimensions of the weakly broken currents: schematically, $\gamma_s=\Delta_s-s-1 \sim \langle \partial \cdot j_s|\partial \cdot j_s\rangle/\langle j_s|j_s\rangle$. Because the right-hand side of (\ref{non-cons-intro}) contains no single-trace operators, it follows that the anomalous dimensions vanish at planar level, and the leading term is of order $1/N$:
\begin{equation}
\label{gamma-s}
\Delta_s = s+1 +\frac{\gamma^{(1)}(s,\lambda)}{N}+\frac{\gamma^{(2)}(s,\lambda)}{N^2}+\ldots \,.
\end{equation}

In this paper, we compute the term of order $1/N$ in the anomalous dimensions (\ref{gamma-s}) for all $s\geq 1$ operators, in both fermionic and bosonic CS vector models, and to all orders in $\lambda$. As described in section \ref{general_analysis}, using the slightly broken higher-spin symmetry, one can show that the anomalous dimensions, or equivalently the twists $\tau_s=\Delta_s-s$, of the HS operators in the bosonic and fermionic CS-vector models must take the form:
\begin{equation}
\label{anom-tau}
\tau_s -1 =\frac{1}{\tilde{N}}\left( a_s  \frac{\tilde{\lambda}^2}{1+\tilde{\lambda}^2} + b_s  \frac{\tilde{\lambda}^2}{(1+\tilde{\lambda}^2)^2}\right)+O(\frac{1}{N^2})\,,
\end{equation}
where $\tilde N$ and $\tilde \lambda$ are the parameters introduced in the analysis of \cite{MZ}, as reviewed in section \ref{CS-actions} and \ref{hs} below. The spin-dependent coefficients $a_s$ and $b_s$ can be determined by computing the 2-point function of the operator appearing in the non-conservation equation (\ref{non-cons-intro}), neglecting the triple-trace term which does not affect the anomalous dimensions to this order. In section \ref{hs} we constrain the divergence of the HS currents using conformal invariance alone, up to some spin-dependent numerical coefficients, and in section \ref{sec:classical} we use the classical equations of motion to calculate the divergence explicitly and fully fix the structure of the double-trace part of (\ref{non-cons-intro}). A priori, the values of $a_s$ and $b_s$ may be different for the fermionic and bosonic theory. However, in our calculations below, we find that they are identical for both theories, and, in the case of $U(N)$ gauge group, they are given by
\begin{eqnarray}
	a_s & = & \begin{cases} \frac{16}{3\pi^2} \frac{s-2}{2s-1}\,, & \text{for even $s$}\,, \\
 \frac{32}{3\pi^2} \frac{s^2-1}{4s^2-1}\,, &  \text{ for odd $s$}\,,\end{cases} \\
	\label{bs-eq-intro}
 b_s & = & \begin{cases}
 \frac{2}{3 \pi
   ^2} \left(3 g(s)+\frac{-38 s^4+24 s^3+34 s^2-24
   s-32}{4
   s^4-5 s^2+1}\right)\,, & \text{for even $s$}\,, \\
    \frac{2}{3 \pi ^2} \left(3
   g(s)+\frac{20-38 s^2}{4
   s^2-1}\right)\,, & \text{ for odd $s$}\,,
\end{cases}
\end{eqnarray}
with \begin{equation}
  g(s)=  \sum_{n=1}^s \frac{1}{n-1/2}= \gamma -\psi(s)+2\psi(2s) = H_{s-1/2}+2 \log(2) \,,
\end{equation}
where $\psi(x)$ is the digamma function, and $H_n$ the Harmonic number. In section \ref{sec:classical}, we also present the results for Chern-Simons theories based on $O(N)$ gauge group, which give slightly different coefficients that are reported in eq. (\ref{as-bs-ON}). As a consistency check, note that the anomalous dimensions vanish for $s=1$ and $s=2$, as expected. 

While the functions $\tilde N$ and $\tilde \lambda$ are not fixed by the weakly broken HS symmetry analysis, they can be fixed by an explicit calculation of 2-point and 3-point functions, and they were found to be \cite{Aharony:2012nh, GurAri:2012is}
\begin{equation}
\tilde N = 2N \frac{\sin(\pi \lambda)}{\pi \lambda}\,,\qquad 
\tilde \lambda = \tan(\frac{\pi \lambda}{2})\,,
\label{tN-tlam}
\end{equation} 
in both CS-scalar and CS-fermion theories, in terms of the respective $N$ and $\lambda$. Using these into (\ref{anom-tau}), the anomalous dimensions take the form
\begin{equation}
\label{anom-Nlam}
\tau_s-1 = \frac{\pi \lambda}{2N \sin(\pi \lambda)}\left(a_s  \sin^2(\frac{\pi\lambda}{2}) +\frac{b_s}{4} \sin^2(\pi\lambda)\right)\,.
\end{equation}
As an independent check of this result, in section \ref{pert} we also perform a direct Feynman diagram calculation in the CS-fermion model, 
from which we find the same values of the $a_s$ and $b_s$ coefficients. 

Note that due to the harmonic sum in (\ref{bs-eq-intro}), we have $b_s \simeq \frac{2}{\pi^2}\log s$ for $s\gg 1$, while $a_s$ is constant at large $s$, 
and so the large spin behavior of the anomalous dimensions is 
\begin{equation}
\tau_s -1 \simeq  \frac{1}{\tilde N}   \frac{\tilde{\lambda}^2}{(1+\tilde{\lambda}^2)^2} \frac{2}{\pi^2}\log s  = \frac{\lambda \sin(\pi \lambda)}{4\pi N}\log s\,.
\end{equation}
This logarithmic behavior is a hallmark of gauge theory, and is expected from general arguments \cite{Alday:2007mf}, see also the recent bootstrap analysis in \cite{Alday:2015ota}. 
The coefficient $f(\lambda)=\frac{\lambda \sin(\pi \lambda)}{4\pi N}$ of $\log s$ may be interpreted as the ``cusp anomalous dimension" of the model; it would be interesting to see if it can be reproduced by computing the expectation value of a Wilson loop with a light-like cusp. 

The result (\ref{anom-Nlam}) applies to the ``regular" CS-fermion and CS-scalar models. In the critical models, a calculation using the classical 
equations of motion, extended to all orders in $\lambda$ by using the results of \cite{MZ}, yields
\begin{equation}
\label{anom-crit}
\tau_s^{\rm crit.} -1 =\frac{1}{\tilde{N}}\left(a_s  \frac{1}{1+\tilde{\lambda}^2} + b_s  \frac{\tilde{\lambda}^2}{(1+\tilde{\lambda}^2)^2}\right)
= \frac{\pi \lambda}{2N \sin(\pi \lambda)}\left(a_s  \cos^2(\frac{\pi\lambda}{2}) +\frac{b_s}{4} \sin^2(\pi\lambda)\right)\,,
\end{equation}
for both the critical CS-scalar and critical CS-fermion theory. In particular, at $\lambda=0$, we recover the anomalous dimensions 
in the usual (Wilson-Fisher) critical $O(N)$ model \cite{Lang:1992zw, Giombi:2016hkj, Skvortsov:2015pea, Hikida:2016wqj} 
and critical GN model \cite{Muta:1976js, Hikida:2016cla}, which happen to coincide in $3d$
\begin{equation}
\gamma_s^{\rm W.F.} = \gamma_s^{\rm GN} = \frac{1}{2N}a_s= \begin{cases} \frac{8}{3N\pi^2} \frac{s-2}{2s-1}\,, & \text{for even $s$}\,, \\
 \frac{16}{3N\pi^2} \frac{s^2-1}{4s^2-1}\,, &  \text{ for odd $s$}\,.\end{cases}
\end{equation}
Note that the same anomalous dimensions arise in the strong coupling limit, $\lambda\rightarrow 1$ ($\tilde\lambda\rightarrow\infty$) 
of the regular CS-fermion and CS-scalar result (\ref{anom-Nlam}). More precisely, in this limit one finds
\begin{equation}
\tau_s-1 \stackrel{\lambda\rightarrow 1}{\simeq} \frac{1}{2(k-N)} a_s\,,
\end{equation}
which are the anomalous dimensions in the $U(k-N)$ critical Wilson-Fisher or Gross-Neveu model. This is a manifestation of the 
``3d bosonization" duality \cite{Giombi:2011kc, MZ, Aharony:2012nh} which conjecturally relates the critical/regular CS-scalar theory 
to the regular/critical CS-fermion theory. The precise form of the duality was spelled out in \cite{Aharony:2012nh, Aharony:2015mjs}, 
and reads\footnote{Versions of this duality map involving the $SU(N)$ 
gauge group were also recently proposed in \cite{Aharony:2015mjs}, where the mapping of baryon and monopole operators was discussed (see also 
\cite{Radicevic:2015yla}).}
\begin{equation}
U(N)_{k-1/2} ~~{\rm CS-Fermion}\quad \Leftrightarrow \quad U(|k|-N)_{-k} ~~ {\rm Critical~CS-Scalar}\,,
\label{lev-rank-1}
\end{equation}
and a similar duality relating the regular CS-scalar to the critical CS-fermion.\footnote{In this case, 
the duality at large $N$ also entails a mapping \cite{GurAri:2012is} between 
the additional marginal couplings $g_6 (\bar\psi\psi)^3$ and $\lambda_6 (\phi^* \phi)^3$ in these models.} So far we have assumed 
that $k$ is the CS level defined in the dimensional reduction scheme \cite{Chen:1992ee}, where no one-loop renormalization of the level occurs. 
To write the duality in a more familiar form, it is useful to express it in terms of $\kappa = k-{\rm sign}(k)N$; this is 
the definition of the CS-level that arises when the theory is regularized with a Yang-Mills term in the UV.\footnote{This definition of $\kappa$ agrees with the level of the WZW theory dual to the CS theory.} In terms of this, the duality reads
\begin{equation}
U(N)_{\kappa-1/2} ~~{\rm CS-Fermion}\quad \Leftrightarrow \quad U(|\kappa|)_{-N} ~~ {\rm Critical~CS-Scalar}\,,
\label{lev-rank-2}
\end{equation}
and it can be recognized as a generalization of level-rank duality in pure CS theory \cite{Naculich:1990pa, Mlawer:1990uv, Camperi:1990dk}. 
Several non-trivial tests of the duality have been obtained in the large $N$ 't~Hooft limit
\cite{Aharony:2012nh,GurAri:2012is,Aharony:2012ns,Jain:2013py,Takimi:2013zca,Jain:2013gza,Jain:2014nza,Moshe:2014bja,Bedhotiya:2015uga,Gur-Ari:2015pca,Geracie:2015drf,Minwalla:2015sca, Yokoyama:2016sbx}. 
If we denote by $N_b$ and $\lambda_b$ the rank and coupling in the critical CS-scalar theory, 
and by $N_f$, $\lambda_f$ the ones in the CS-fermion theory, in the large $N$ limit (where we can neglect the half-integer shift of the level on the fermionic 
side), the duality implies the map
\begin{equation}
\lambda_b = \lambda_f-{\rm sign}(\lambda_f)\,,\qquad \frac{N_b}{|\lambda_b|}= \frac{N_f}{|\lambda_f|}\,,
\label{dualMap}
\end{equation}
or equivalently $\tilde N_b = \tilde N_f$, $\tilde\lambda_b^2 = 1/\tilde\lambda_f^2$. 
Comparing (\ref{anom-Nlam}) and (\ref{anom-crit}), we see that the anomalous 
dimensions are indeed mapped into each other under the duality. Furthermore, by our explicit calculation using the classical equations of motion in section 
\ref{sec:classical}, we will verify that the non-conservation equations (\ref{non-cons-intro}) in the dual theories 
correctly map into each other, including the normalization factors. 

The ``3d bosonization" (\ref{lev-rank-1}) may also be regarded as a non-supersymmetric version of the supersymmetric dualities 
\cite{Giveon:2008zn, Benini:2011mf}, which are well established at finite $N$ and $k$. 
Therefore, it is plausible that the bose/fermi duality (\ref{lev-rank-1}) holds away from the 
large $N$ limit. For small $N$ and $k$, (\ref{lev-rank-1}) and related dualities may 
have interesting applications in condensed matter physics, see for example 
\cite{Karch:2016sxi, Murugan:2016zal, Seiberg:2016gmd, Kachru:2016rui,Radicevic:2016wqn,Hsin:2016blu} for recent closely related work. 
While exact results at finite $N$ and $k$ are hard to obtain, it would be interesting 
to see if the subleading terms in the large $N$ expansion of the anomalous dimensions (or other quantities 
such as the thermal free energy) may be also computed 
for finite $\lambda$, and whether they agree with the duality. Note that the half-integer shift in the CS-fermion level can play a non-trivial 
role in this case. As a first step towards determining subleading corrections at large $N$, in section \ref{sec:classical} 
we use the classical equations of motion method to fix the terms of order $\lambda^2/N^2$ in the anomalous dimensions of the CS-scalar and CS-fermion models. 
In particular, this result gives the term of order $1/k^2$ in the scaling dimensions of the spin-$s$ operators in the $U(1)_{k}$ CS theory coupled to a fundamental 
fermion.\footnote{In one version of the dualities put forward in \cite{Aharony:2015mjs}, see also \cite{Seiberg:2016gmd}, the $U(1)_{-1/2}$ CS-fermion theory is related 
to the critical $O(2)$ model without CS gauge field. Our result for the anomalous dimensions $\gamma_s$ in the $U(1)_k$ theory to order $1/k^2$ shows 
logarithmic behavior at large $s$. On the other hand, we do not expect logarithmic growth in the critical $O(2)$ model. It is plausible that the 
$\log s$ behavior disappears in the strongly coupled ($k=-1/2$) theory, but it would be interesting to understand this better.}   

Besides encoding the anomalous dimensions of the HS operators, the current non-conservation equation (\ref{non-cons-intro}) can also be used 
to completely fix (including the overall normalization) the parity odd structure in the planar 3-point functions 
of $\langle j_{s_1} j_{s_2} j_{s_3}\rangle$ when the triangular inequality is violated, 
i.e. $s_3> s_1+s_2$; this is the case where the 3-point function breaks the $j_{s_3}$ current conservation. 
In section \ref{three-point-section}, we use our results from the classical divergence calculation to determine explicitly 
all such parity odd 3-point functions. In particular, we derive some recursion relations that can be used to obtain the explicit form of the 
3-point functions for general spins. The parity-odd three-point functions are further analyzed in Appendix \ref{three-point-appendix}, 
with some examples listed for low spins in Appendix \ref{tables}.

An interesting open problem that we do not address in this paper is the calculation of the scaling dimension of the scalar operators $\bar\phi\phi$ or 
$\bar\psi \psi$. It is possible to argue \cite{Giombi:2011kc,Aharony:2011jz} based on the structure of the HS breaking equations (where 
the scalar operators can appear on the right-hand side) that they must have dimensions $\Delta=1+O(1/N)$ or $\Delta=2+O(1/N)$, but 
it is not obvious if the weakly broken HS symmetry can be used to determine the order-$1/N$ correction for finite $\lambda$. A direct 
all-orders diagrammatic calculation may in principle be possible, but it appears to require a currently unavailable ingredient: the ladder diagram of \cite{Bedhotiya:2015uga, Jain:2014nza} for general off-shell external momenta.

Another interesting direction would be to extend the results of this paper to various other related CS-matter theories. As an example, $U(N)\times U(M)$ 
Chern-Simons theories coupled to bi-fundamental matter, also possess a weakly broken HS symmetry when $M/N \ll 1$ 
\cite{Chang:2012kt, Banerjee:2013nca, Gurucharan:2014cva}, and the methods used in this paper should be applicable 
to this class of models. As the non-supersymmetric theories have two independent Chern-Simons levels, the $1/N$ anomalous dimensions here appear to depend on two independent parameters, so it would be interesting to see how these parameters relate to the general analysis of \cite{MZ} (which, in its present form, applies to theories with even spin currents only).  It may be also interesting to consider general CS-vector models \cite{Jain:2013gza} with fundamental boson and fermions on the same side (including 
in particular the supersymmetric theories as a special case). 

Perhaps the most interesting extension of this work would be to calculate the anomalous dimensions of higher spin operators in the $\mathcal N=6$  ABJ theory \cite{Aharony:2008gk}, in the regime $M \ll N$, which has been conjectured to be dual to a particular limit of type IIA string theory. Our results here do not directly carry over to this
case because of the additional matter fields and the presence of the
Chern-Simons coupling for the second gauge field, but we expect a
similar analysis to be possible in principle. We hope to return to
this in future work.

\begin{figure}
\begin{center}
\includegraphics[width=0.7\textwidth]{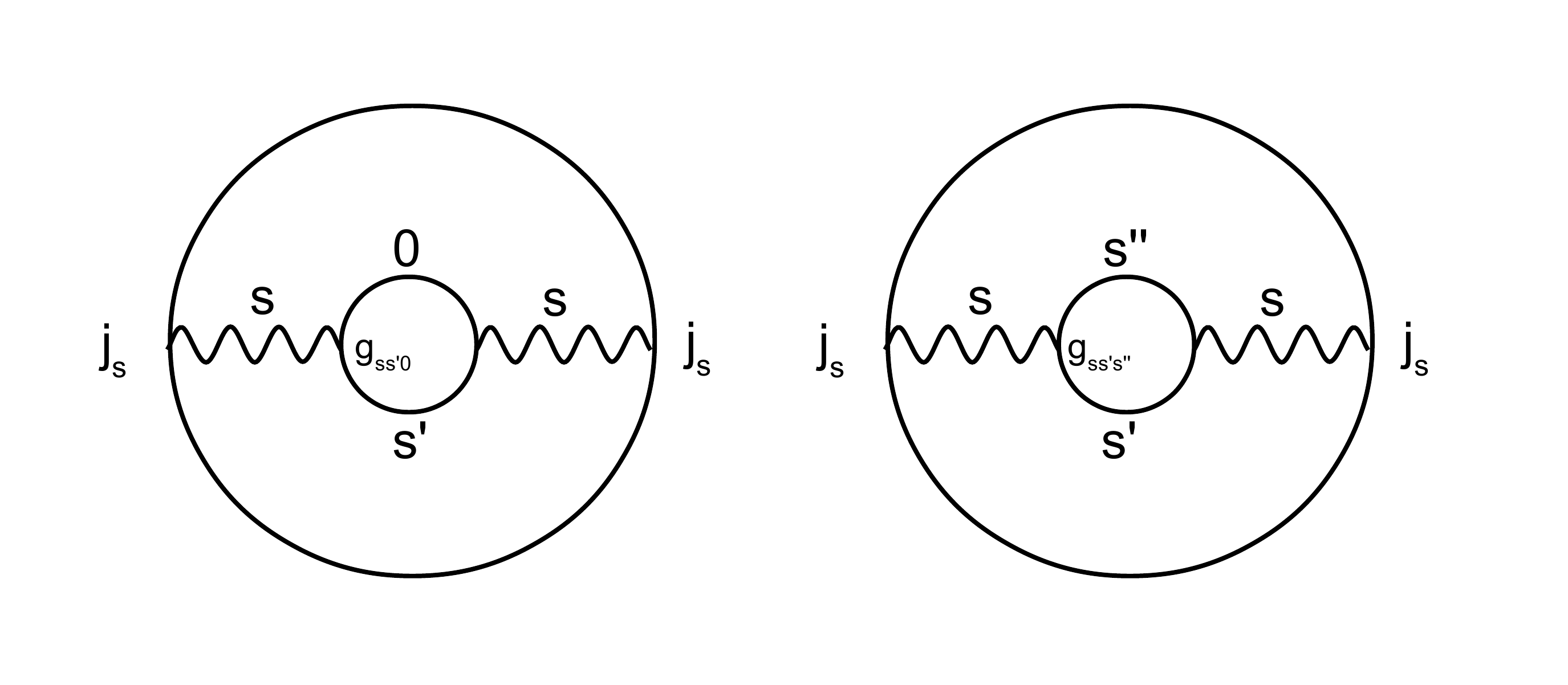}
\end{center}
\label{2pt-bulk}
\vskip -0.5cm
\caption{The one-loop bulk diagrams that are expected to reproduce the $1/N$ term in the anomalous dimensions of the HS currents at the boundary.}
\end{figure}

As mentioned earlier, the weakly broken HS operators should correspond in the dual AdS$_4$ theory to classically massless HS gauge fields that 
acquire masses via loop corrections, through a HS analogue of the Higgs mechanism \cite{Girardello:2002pp}.\footnote{The role of the Higgs field 
is played in this case by a multi-particle state in the bulk which is dual to the operator appearing on the right-hand side 
of (\ref{non-cons-intro}).} It would be interesting to see if the result for the anomalous dimensions (\ref{anom-Nlam}) can be 
reproduced by a one-loop calculation in the parity breaking higher-spin theory, corresponding schematically to the diagrams depicted 
in figure \ref{2pt-bulk}. Note that the coupling constant in the bulk is fixed by the duality to be 
$1/G_N \sim \tilde{N} = 2N\frac{\sin(\pi\lambda)}{\pi \lambda}$, 
and the parity breaking 3-point couplings are expected to depend on the bulk parameter $\theta_0$ as 
$g^{\rm odd}_{ss'0}\sim \sin\theta_0$ and $g^{\rm odd}_{ss's''}\sim \sin(2\theta_0)$ (see e.g. \cite{Giombi:2012ms}), and 
we also have $g^{\rm even-A}_{ss's''}\sim \cos^2(\theta_0)$, $g^{\rm even-B}_{ss's''}\sim \sin^2(\theta_0)$. Therefore we see that if $\theta_0 = \pi\lambda/2$, 
which is required for agreement of the tree-level 3-point functions, the bulk one-loop diagrams would yield the expected coupling dependence 
we found in (\ref{anom-Nlam}). It remains to be seen if the spin-dependent coefficients can be reproduced from the AdS calculation. 

\section{The Chern-Simons vector models}
\label{CS-actions}
The action for the $U(N)$ Chern-Simons theory at level $k$ coupled to a massless fundamental scalar field is given in our conventions by
\begin{equation}
    S=\frac{ik}{4\pi}S_{\rm CS}+\int d^3x \left( D_{\mu}\bar{\phi} D^{\mu}\phi +\frac{\lambda_6}{N^2}  (\bar\phi \phi)^3\right)\,,
\label{CS-boson}
\end{equation}
where
\begin{equation}
S_{\rm CS} =\int d^3x \epsilon^{\mu\nu\rho} \mbox{Tr} (A_{\mu} \partial_\nu A_\rho - \frac{2i}{3} A_\mu A_\nu A_\rho)\,.
\end{equation}
We work in Euclidean signature throughout the paper, and use the conventions $D_\mu\phi = \partial_{\mu}\phi - i A_\mu \phi$, $D_\mu\bar\phi = \partial_{\mu}\bar\phi + i \bar\phi A_{\mu}$, with $A_\mu=A^a_{\mu} T^a$, where $T^a$ are the generators of $U(N)$ in the fundamental representation. One can show that in the large $N$ limit with $\lambda=N/k$ and $\lambda_6$ fixed, the classically marginal coupling $\lambda_6$ is in fact exactly marginal. Hence, in the large $N$ limit the model (\ref{CS-boson}) defines a CFT (provided the scalar mass is suitably tuned to zero) labelled by two marginal parameters $\lambda$, $\lambda_6$ \footnote{Away from the $N\rightarrow \infty$ limit, $\beta_{\lambda_6}(\lambda_6,\lambda)\neq 0$, but one finds fixed points with $\lambda_6=\lambda_6^*(\lambda)$ \cite{Aharony:2011jz}.}. The value of $\lambda_6$ does not affect the anomalous dimensions of the higher-spin operators to the order $1/N$ we consider, and hence we will neglect this coupling in the following.

One may define another bosonic CFT, sometimes referred to as the critical bosonic theory, by adding to (\ref{CS-boson}) a quartic interaction $\frac{\lambda_4}{N}
(\bar\phi \phi)^2$ and flowing to the infrared. Rewriting the quartic coupling with the aid of a Hubbard-Stratonovich auxiliary field $\sigma_b$, the action of the IR CFT may be written as
\begin{equation}
S_{\rm crit} = \frac{ik}{4\pi}S_{\rm CS}+\int d^3x \left( D_{\mu}\bar{\phi} D^{\mu}\phi+\frac{1}{N}\sigma_b \bar\phi \phi \right)\,,
\label{crit-bos}
\end{equation}
where the quadratic term in $\sigma_b$ was dropped, which is appropriate in the IR limit. The factor of $1/N$ was introduced so that the 2-point function of
$\sigma_b$ scales like $N$. Note that the $\phi^6$ term can be dropped since this coupling becomes irrelevant in the IR. This model defines a generalization of the Wilson-Fisher CFT by the addition of the Chern-Simons gauge coupling.

The action of a fundamental massless fermion coupled to the $U(N)$ CS gauge field at level $k$ is given by
\begin{equation}
    S=\frac{ik}{4\pi}S_{\rm CS}+\int d^3x \bar{\psi} \slashed{D} \psi\,,
\label{CS-fermion}
\end{equation}
where we define $\slashed{D}=\gamma^{\mu} D_{\mu}$ and $D_\mu\psi  = \partial_\mu\psi - i A_\mu \psi$. Note that the level $k$ should be half-integer due to the parity anomaly, however this condition will not be important for the large $N$ computations we will perform below.  The action (\ref{CS-fermion}) defines a CFT labeled by $N$ and $\lambda =N/k$, provided the fermion mass term is tuned to zero.

Analogously to the scalar case, one may add to the model (\ref{CS-fermion}) a quartic self-interaction $\frac{g_4}{N}(\bar\psi\psi)^2$. Such theory is expected to have, at least in the large $N$ limit, a non-trivial UV fixed point which is a generalization of the critical 3d Gross-Neveu model. The action describing the UV CFT can be taken to be
\begin{equation}
S_{\rm crit} = \frac{ik}{4\pi}S_{\rm CS}+\int d^3x \left(\bar{\psi} \slashed{D} \psi+\frac{1}{N}\sigma_f \bar\psi \psi \right)\,,
\label{crit-fer}
\end{equation}
where $\sigma_f$ is the auxiliary Hubbard-Stratonovich field, and the quadratic term was dropped as appropriate in the UV limit. At large $N$, the model also possesses an exactly marginal coupling $g_6 (\bar\psi \psi)^3 \sim g_6 \sigma_f^3$. This extra coupling (which is mapped under the bose-fermi duality to the $\lambda_6$ coupling in the CS-boson theory) does not affect the quantities we will compute in this paper, and we will neglect it below.

\subsection{The ``single-trace" operators}

\paragraph{Free theories} Let us first review the spectrum of ``single-trace" operators in the free bosonic and fermionic $U(N)$ vector models. In the scalar model, the spectrum
consists of a scalar operator
\begin{equation}
j_0 = \bar\phi \phi
\end{equation}
with scaling dimension $\Delta=1$, and a tower of exactly conserved currents $j_s \sim \bar\phi \partial^s \phi$ of all integer spins. To give the explicit
form of these currents, it is convenient to introduce an auxiliary null vector $z^{\mu}$, $z^{\mu} z_{\mu}=0$, and define the index-free operators
\begin{equation}
j_s(x,z) = j_{\mu_1\cdots \mu_s}z^{\mu_1}\cdots z^{\mu_s}\,.
\end{equation}
A generating function $J_b(x,z) = \sum_{s=0}^{\infty} j_s(x,z)$ of the higher-spin operators in the scalar theory is given by \cite{Giombi:2009wh}
\begin{equation}
\begin{aligned}
&J_{\rm b} = \bar\phi(x) f_{\rm b}( z \cdot \overleftarrow{\partial},z \cdot \overrightarrow{\partial})\phi(x)
= f_{\rm b}(\hat \partial_1,\hat\partial_2)\bar\phi(x_1)\phi(x_2)|_{x_1,x_2\rightarrow x}\,,\\
&f_{\rm b}(u,v) = e^{u-v}\cos(2\sqrt{u v})\,.
\label{HS-sc}
\end{aligned}
\end{equation}
In the first line, we have introduced a bilocal notation which will be useful below, and we defined the shorthand $\hat\partial \equiv z\cdot \partial$. One
may restore the explicit indices on the currents by acting with the differential operator in $z$-space
\cite{Dobrev:1975ru, Craigie:1983fb, Belitsky:2007jp, Costa:2011mg}
\be
\label{Dz}
D^{\mu}_z\equiv \frac{1}{2}\partial_{z_\mu} +z^\nu \partial_{z_\nu} \partial_{z_\mu} - \frac{1}{2} z^\mu \partial_{z_\nu} \partial_{z_\nu}.
\ee
For instance, to compute the divergence of the current $j_s$, one can evaluate
$\partial^{\mu} D_{\mu}^z j_s(x,z) \propto \partial^{\mu}j_{\mu \mu_2\cdots\mu_s}z^{\mu_2}\cdots z^{\mu_s}$. Using the free equation of motion $\partial^2\phi=0$, one can explicitly check that the currents in (\ref{HS-sc}) are conserved. Indeed,
the condition $\partial^{\mu}D_{\mu}^z J_{\rm b} = 0$ turns into the differential equation
\begin{equation}
\left(\frac{1}{2}(\partial_u+\partial_v) + u \partial_{u}^2  + v \partial_{v}^2 \right) f_{\rm b}(u,v)=0\,,
\end{equation}
which is seen to be satisfied by the generating function given above. Expanding (\ref{HS-sc}) in powers of $z$, one may also deduce the following explicit
expression for the currents
\begin{equation}
j^{\rm b}_s(x,z) = \sum_{k=0}^s  \frac{(-1)^{k+s}}{s!}\begin{pmatrix}2s\\2k \end{pmatrix}
\hat\partial_1^k \hat\partial_2^{s-k} \bar\phi(x_1)\phi(x_2)|_{x_1,x_2\rightarrow x}\,.
\end{equation}
Using the free scalar propagator
\begin{equation}
\langle \bar\phi(x)\phi(0)\rangle = \frac{1}{4\pi|x|}\,,
\end{equation}
it is straightforward to derive the 2-point function normalization of the higher-spin operators in the free scalar theory. One finds
\begin{equation}
\begin{aligned}
&\langle j^{\rm b}_s(x,z) j^{\rm b}_s(0,z)\rangle =N n_s \frac{(z\cdot x)^{2s}}{(x^2)^{2s+1}}\,,\\
&n_s = \frac{2^{4 s-5} \Gamma \left(s+\frac{1}{2}\right)}{\pi^{5/2} s!} \,.
\label{js-norm-sc}
\end{aligned}
\end{equation}
Similarly, for the $\Delta=1$ scalar we have
\begin{equation}
\langle j_0(x) j_0(0)\rangle = \frac{N}{16\pi^2 x^2}\equiv \frac{N n_0}{x^2}\,.
\label{j0-norm}
\end{equation}

In the free fermionic $U(N)$ vector model, the single-trace operators consist of the parity odd scalar
\begin{equation}
\tilde{j}_0 = \bar\psi\psi
\end{equation}
with $\Delta=2$, and the conserved higher-spin currents $j_s \sim \bar\psi \gamma \partial^{s-1}\psi$, given
explicitly by the generating function \cite{Giombi:2011kc}
\begin{equation}
\begin{aligned}
&J_{\rm f}= \bar\psi(x) z\cdot \gamma f_{\rm f}( z \cdot \overleftarrow{\partial},z \cdot \overrightarrow{\partial})\psi(x)
= f_{\rm f}(\hat \partial_1,\hat\partial_2)\bar\psi(x_1)\hat\gamma\psi(x_2)|_{x_1,x_2\rightarrow x}\,,\\
&f_{\rm f}(u,v) = e^{u-v}\frac{\sin(2\sqrt{u v})}{2\sqrt{uv}}\,.
\label{HS-fer}
\end{aligned}
\end{equation}
To check that these currents are conserved when $\psi$ obeys the free equation of motion, one can verify that
$\partial^{\mu}D_{\mu}^z J_{\rm f}(x,z)=0$. This yields
\begin{equation}
\left(\frac{3}{2}(\partial_u+\partial_v) + u \partial_{u}^2  + v \partial_{v}^2 \right) f_{\rm f}(u,v)=0\,,
\end{equation}
which is satisfied by the generating function in (\ref{HS-fer}). By expanding in powers of $z$, one can also derive the following explicit form
\begin{equation}
j^{\rm f}_s(x,z) = \sum_{k=0}^{s-1}  \frac{(-1)^{k+s+1}}{2s!}\begin{pmatrix}2s\\2k +1\end{pmatrix}
\hat\partial_1^k \hat\partial_2^{s-k-1} \bar\psi(x_1)\hat\gamma\psi(x_2)|_{x_1,x_2\rightarrow x}\,.
\label{jsf}
\end{equation}
Using the free fermion propagator
\begin{equation}
\langle \psi(x)\bar\psi(0)\rangle = \frac{1}{4\pi}\frac{\slashed{x}}{x^3}\,,
\end{equation}
one finds that the currents in the free fermion theory (\ref{HS-fer}), (\ref{jsf}) have exactly the same 2-point normalization
as the scalar ones
\begin{equation}
\begin{aligned}
&\langle j^{\rm f}_s(x,z) j^{\rm f}_s(0,z)\rangle =N n_s \frac{(z\cdot x)^{2s}}{(x^2)^{2s+1}}\,,\qquad
n_s = \frac{2^{4 s-5} \Gamma \left(s+\frac{1}{2}\right)}{\pi^{5/2} s!} \,.
\label{js-norm-fer}
\end{aligned}
\end{equation}
For the parity odd scalar operator, one finds
\begin{equation}
\langle \tilde{j}_0(x) \tilde{j}_0(0)\rangle = \frac{N}{8\pi^2 x^4}\equiv \frac{N \tilde{n}_0}{x^4}\,.
\label{tj0-norm}
\end{equation}

In the calculations below, we will sometimes find it convenient to introduce explicit light-cone coordinates, with metric
\begin{equation}
ds^2=2 dx^+ dx^-+dx_3^2\,.
\end{equation}
When we do this, we will take the auxiliary null vector to be $z^{\mu}=\delta^{\mu}_{-}$, and so $j_s(x,z)= j^s_{--\cdots-}$, $\hat\partial = \partial_-$ and $z\cdot x=x_-$.

\paragraph{Interacting theories}
When the Chern-Simons coupling is turned on, the higher-spin operators defined above should be made gauge invariant by replacing derivatives
with covariant derivatives. The currents in the bosonic theories are then
\begin{equation}
\begin{aligned}
&J_{\rm b} = \bar\phi(x) f_{\rm b}(\overleftarrow{\hat D}, \overrightarrow{\hat D})\phi(x)=\sum_{s=0}^{\infty}j^{\rm b}_s(x,z)\,, \\
&j^{\rm b}_s(x,z) = \sum_{k=0}^s  \frac{(-1)^{k+s}}{s!}\begin{pmatrix}2s\\2k \end{pmatrix}
\hat D_1^k \hat D_2^{s-k} \bar\phi(x_1)\phi(x_2)|_{x_1,x_2\rightarrow x}\,,
\label{Jb-CS}
\end{aligned}
\end{equation}
where $\hat D = z^{\mu}D_{\mu}$, and recall $\hat D \phi = \hat\partial\phi-i \hat A \phi$, $\hat D\bar\phi = \hat\partial \bar\phi +i \bar\phi \hat A$.
Similarly, in the fermionic theories one has
\begin{equation}
\begin{aligned}
&J_{\rm f} = \bar\psi(x) \hat\gamma f_{\rm f}(\overleftarrow{\hat D}, \overrightarrow{\hat D})\psi(x)=\sum_{s=1}^{\infty}j^{\rm f}_s(x,z)\,, \\
&j^{\rm f}_s(x,z) = \sum_{k=0}^s  \frac{(-1)^{k+s+1}}{2s!}\begin{pmatrix}2s\\2k+1 \end{pmatrix}
\hat D_1^k \hat D_2^{s-k} \bar\psi(x_1)\hat\gamma \psi(x_2)|_{x_1,x_2\rightarrow x}\,.
\label{Jf-CS}
\end{aligned}
\end{equation}
Note that contracting with the null vector $z^{\mu}$ automatically projects the currents onto their symmetric traceless part. The higher-spin operators
above, together with the scalars $j_0=\bar\phi \phi$ and $\tilde j_0=\bar\psi \psi$, exhaust the single-trace spectrum in the interacting theories as well
\cite{Giombi:2011kc, Aharony:2011jz}. Note that the CS equation of motion
$\frac{k}{4\pi}\epsilon^{\mu\nu\rho}(F_{\nu\rho})^{i}_{\ j}=(J^{\mu})^i_{\ j}$, where $(J_{\mu})^{i}_{\ j}$ is the $U(N)$ current, implies that
naive single-trace operators obtained by inserting factors of the field strength inside matter bilinears are in fact multi-trace.

In the interacting theory the higher-spin currents are no longer conserved, however the breaking is small at large $N$ and implies that anomalous dimensions
are generated starting at order $1/N$.
The 2-point and 3-point functions of the bilinear operators can be fixed in the planar limit and to all orders in $\lambda$ by using the weakly broken
higher-spin symmetry \cite{MZ} and explicit computations for low spins \cite{Aharony:2012nh,GurAri:2012is}.

In the CS-boson theory, one finds for the exact planar 2-point functions \cite{Aharony:2012nh}
\begin{equation}
\begin{aligned}
&\langle j^{\rm b}_s(x,z) j^{\rm b}_s(0,z)\rangle =N\frac{\sin(\pi\lambda)}{\pi\lambda}n_s \frac{(z\cdot x)^{2s}}{(x^2)^{2s+1}}\,, \\
&\langle j_0(x) j_0(0)\rangle = \frac{2N\tan(\frac{\pi\lambda}{2})}{\pi\lambda}\frac{n_0}{x^2}\,.
\end{aligned}
\end{equation}
In terms of the parameters $\tilde N$ and $\tilde \lambda$ introduced in the analysis of \cite{MZ}, these read\footnote{Note that the scalar
operator $j_0 =\bar\phi\phi$ has a different normalization from the one chosen in \cite{MZ}. They are related by $j_0^{\rm MZ} = j_0/(1+\tilde\lambda^2)$.
Similarly, in the fermionic theory we define $\tilde{j}_0 = \bar\psi\psi$, $\tilde{j}_0^{\rm MZ} =  \tilde{j}_0/(1+\tilde\lambda^2)$.}
\begin{equation}
\begin{aligned}
&\langle j^{\rm b}_s(x,z) j^{\rm b}_s(0,z)\rangle =\tilde N \langle j_s(x,z) j_s(0,z) \rangle_{\rm sc}\,,\\
&\langle j_0(x) j_0(0)\rangle = \tilde N(1+\tilde\lambda^2) \langle j_0(x) j_0(0) \rangle_{\rm sc}\,,
\label{2pt-CSbos}
\end{aligned}
\end{equation}
where the correlators on the right-hand side refer to the theory of a single real free scalar, and we used \cite{Aharony:2012nh}
\begin{equation}
\tilde N = 2N\frac{\sin(\pi\lambda)}{\pi\lambda}\,,\qquad \tilde\lambda = \tan(\frac{\pi\lambda}{2})\,.
\label{tNtlam}
\end{equation}
The 3-point functions of operators of non-zero spin are fixed to be
\begin{equation}
\langle j^{\rm b}_{s_1} j^{\rm b}_{s_2}j^{\rm b}_{s_3}\rangle = \tilde N
\left[\frac{1}{1+\tilde\lambda^2}\langle j_{s_1} j_{s_2}j_{s_3}\rangle_{\rm sc}
+\frac{\tilde\lambda^2}{1+\tilde\lambda^2}\langle j_{s_1} j_{s_2}j_{s_3}\rangle_{\rm fer}
+\frac{\tilde\lambda}{1+\tilde\lambda^2}\langle j_{s_1} j_{s_2}j_{s_3}\rangle_{\rm odd}\right]\,,
\label{CSbos-sss}
\end{equation}
where the suffix `sc' and `fer' refer to the correlators in the (real) free scalar and free fermion theories, and the `odd' term
is a structure which breaks parity. It also breaks current conservation when $s_1,s_2,s_3$ do not satisfy the triangular inequality,
as will be explained below. When one of the operators is the scalar $j_0 =\bar\phi \phi$, the 3-point functions read
\begin{equation}
\langle j^{\rm b}_{s_1} j^{\rm b}_{s_2}j_0 \rangle = \tilde N
\left[\langle j_{s_1} j_{s_2}j_{0}\rangle_{\rm sc}
+\tilde\lambda \langle j_{s_1} j_{s_2}j_{0}\rangle_{\rm odd}\right]\,.
\label{CSbos-ss0}
\end{equation}
Here $\langle j_{s_1} j_{s_2}j_{0}\rangle_{\rm odd}$ is a parity odd tensor structure that breaks the spin $s_1$ current conservation when $s_1>s_2$.
Similarly one can write down the expression for correlators involving two or three scalar operators: these are completely fixed by conformal invariance
up to the overall constant, and do not play a role in the analysis of the higher-spin anomalous dimensions to order $1/N$.

In the CS-fermion theory (\ref{CS-fermion}), one finds the analogous results \cite{GurAri:2012is}
\begin{equation}
\begin{aligned}
&\langle j^{\rm f}_s(x,z) j^{\rm f}_s(0,z)\rangle =\tilde N  \frac{n_s(z\cdot x)^{2s}}{2(x^2)^{2s+1}}
=\tilde N \langle j_s(x,z) j_s(0,z) \rangle_{\rm fer}\,,\\
&\langle \tilde{j}_0(x) \tilde{j}_0(0)\rangle = \tilde N(1+\tilde\lambda^2) \frac{\tilde{n}_0}{2x^4}
=\tilde N(1+\tilde\lambda^2) \langle \tilde{j}_0(x) \tilde{j}_0(0) \rangle_{\rm fer}\,,
\label{2pt-CSfer}
\end{aligned}
\end{equation}
where the subscript `fer' indicates correlators in the free theory of a single real fermion. The parameters $\tilde N$ and $\tilde\lambda$ are given in terms of $N$, $\lambda$ by the same expressions as in (\ref{tNtlam}). The 3-point
functions are
\begin{equation}
\langle j^{\rm f}_{s_1} j^{\rm f}_{s_2}j^{\rm f}_{s_3}\rangle = \tilde N
\left[\frac{1}{1+\tilde\lambda^2}\langle j_{s_1} j_{s_2}j_{s_3}\rangle_{\rm fer}
+\frac{\tilde\lambda^2}{1+\tilde\lambda^2}\langle j_{s_1} j_{s_2}j_{s_3}\rangle_{\rm sc}
+\frac{\tilde\lambda}{1+\tilde\lambda^2}\langle j_{s_1} j_{s_2}j_{s_3}\rangle_{\rm odd}\right]\,,
\label{CSfer-sss}
\end{equation}
and
\begin{equation}
\langle j^{\rm f}_{s_1} j^{\rm f}_{s_2}\tilde{j}_0 \rangle = \tilde N
\left[\langle j_{s_1} j_{s_2}\tilde{j}_{0}\rangle_{\rm fer}
+\tilde\lambda \langle j_{s_1} j_{s_2}\tilde{j}_{0}\rangle_{\rm odd}\right]\,.
\label{CSfer-ss0}
\end{equation}
The `odd' structure in the above equation breaks current conservation on $j_{s_1}$ when $s_1>s_2$. It breaks parity, but note that since
$\tilde{j}_0=\bar\psi\psi$ is parity odd, this tensor structure is actually parity even.

Let us now discuss the critical models defined by (\ref{crit-bos}) and (\ref{crit-fer}). In the scalar theory, the auxiliary field $\sigma_b$
replaces the scalar operator $\bar\phi \phi$, and in the IR it behaves as a scalar operator with scaling dimension $\Delta=2+O(1/N)$. To leading order
at large $N$, its two-point function is essentially the inverse of the $\bar\phi \phi$ 2-point function (in momentum space), and reads
\begin{equation}
\langle \sigma_b(x)\sigma_b(0)\rangle = N \frac{4\pi\lambda}{\tan(\frac{\pi\lambda}{2})} \frac{1}{\pi^2 x^4}\,.
\end{equation}
Note that this result is valid to all orders in $\lambda$. Defining the operator $\tilde j_{0}^{\rm crit.bos.} =\sigma_b/(4\pi\lambda)$, one finds
\begin{equation}
\langle \tilde{j}_0^{\rm crit.bos.}(x)\tilde{j}_0^{\rm crit.bos.}(0)\rangle =
 \frac{N}{4\pi\lambda\tan(\frac{\pi\lambda}{2})} \frac{1}{\pi^2 x^4}=\tilde N(1+\frac{1}{\tilde{\lambda}^2}) \frac{\tilde{n}_0}{2x^4}\,.
\end{equation}
We see that this 2-point function precisely matches the $\tilde{j}_0$ 2-point function in the fermionic theory, eq. (\ref{2pt-CSfer}), under the 
duality map (\ref{dualMap}). To leading order at large $N$,the 2-point and 3-point functions involving operators with spin are unchanged in the critical 
theory compared to the CS-boson theory, and the
agreement with the duality follows by comparing (\ref{CSbos-sss}) and (\ref{CSfer-sss}). The 3-point functions involving one (or more)
scalars $\sigma_b$ can be obtained from the ones in the CS-boson theory by attaching a $\sigma_b$ line to the every scalar operator $\bar\phi\phi$, using
the vertex in (\ref{crit-bos}). In terms of $\tilde j_{0}^{\rm crit.bos.} =\sigma_b/(4\pi\lambda)$, the corresponding 3-point functions are related
to those of the CS-fermion theory (\ref{CSfer-ss0}) by the duality map (\ref{dualMap}). Note that the tensor structure
$\langle j_{s_1}j_{s_2}\tilde{j}_0\rangle_{\rm odd}$ in (\ref{CSfer-ss0}) corresponds to the correlators of the critical $O(N)$ model (Wilson-Fisher), which is recovered
in the $\tilde\lambda_f\rightarrow \infty$ limit of the CS-fermion model (or $\tilde\lambda_b\rightarrow 0$ limit of the critical CS-boson model).

The discussion of the critical fermion model (\ref{crit-fer}) goes similarly. The auxiliary field $\sigma_f$ becomes a scalar primary with dimension $\Delta=1+O(1/N)$
in the UV, and its 2-point function can be computed to be
\begin{equation}
\langle \sigma_f(x)\sigma_f(0)\rangle = N \frac{2\pi\lambda}{\tan(\frac{\pi\lambda}{2})} \frac{1}{\pi^2 x^2}\,.
\end{equation}
The duality with the CS-boson model can be verified by defining the operator $j_0^{\rm crit.fer.} = \sigma_f/(4\pi\lambda)$, which has the
2-point function
\begin{equation}
\langle j_0^{\rm crit.fer.}(x) j_0^{\rm crit.fer.}(0)\rangle =
 \frac{N}{8\pi\lambda\tan(\frac{\pi\lambda}{2})} \frac{1}{\pi^2 x^2}=\tilde N(1+\frac{1}{\tilde{\lambda}^2}) \frac{n_0}{2x^2}\,.
\end{equation}
This matches the CS-boson 2-point function (\ref{2pt-CSbos}) under the duality map (\ref{dualMap}). Similarly, the 3-point functions involving a scalar
can be seen to map to those of the CS-boson theory. The tensor structure in (\ref{CSbos-ss0}) which breaks current conservation corresponds to the
correlators of the critical Gross-Neveu model, which is recovered in the limit $\tilde\lambda_f=0$ of the critical CS-fermion (or $\tilde\lambda_b\rightarrow
\infty$ in the CS-boson model). Note that in this case there is an additional marginal parameter $g_6$ on both sides of the duality, as discussed earlier, and
a corresponding duality map \cite{GurAri:2012is}. We will neglect this coupling throughout the paper.

\section{Analysis based on Slightly Broken Higher-Spin Symmetry}
\label{hs}
The theories we study have a tower of single-trace primary spin-$s$ operators $j_s$ which have scaling dimension $\Delta=s+1+O(1/N)$ and are nearly conserved currents \cite{Giombi:2011kc, Aharony:2011jz,MZ}. Following the terminology introduced in \cite{MZ}, we call ``quasi-boson" theory the CFT whose single trace spectrum include, in addition to the spin-$s$ operators, a scalar $j_0$ with $\Delta=1+O(1/N)$; and ``quasi-fermion" theory the CFT with a scalar $\tilde{j}_0$ of dimension
$\Delta=2+O(1/N)$. The ``regular" CS-boson theory or critical CS-fermion theory fall in the quasi-boson class, while the regular CS-fermion or critical CS-boson fall in the quasi-fermion class.

In \cite{MZ}, the quasi-bosonic and quasi-fermionic theories are defined in terms of two parameters: $\tilde{\lambda}$ and $\tilde{N}$. (In the quasi-bosonic theory there is an additional parameter $\tilde{\lambda}_6$ which we ignore here.) The parameter $\tilde N$ can be defined via the normalization of the spin 2 operator (the stress-tensor) two-point function, while $\tilde\lambda$ is defined via the spin 4 anomalous conservation relation:
\begin{equation}
\label{dj4-MZ}
	\partial \cdot j_4 \sim \frac{\tilde{\lambda}}{\tilde{N}} \left( \partial_- \tilde{j}_0^{\rm MZ} j_2 -\frac{2}{5} \tilde{j}_0^{\rm MZ} \partial_- j_2 \right)
\end{equation}
in the quasi-fermion case, and similarly in the quasi-boson case. Here $\sim$ denotes equality up to a $\tilde{\lambda}$-independent numerical coefficient, and $j_0^{\rm MZ}$ denotes the scalar in the normalizations used in \cite{MZ}, which differ from ours by $\tilde{j}_0^{\rm MZ} = \tilde{j}_0/(1+\tilde\lambda^2)$. With $\tilde{\lambda}$ so defined, \cite{MZ} derive expressions for all two-point functions and three-point functions of single-trace primary operators $j_s$.
\subsection{General form of current non-conservation}
\label{general_analysis}
To derive the general expression for anomalous dimensions of spin $s$ currents, we need an expression for the divergence of $j_s$. As argued in \cite{Giombi:2011kc, Aharony:2011jz, MZ} the divergence of $j_s$ (for $s>0$) takes the following form:
\begin{eqnarray}
\partial \cdot j_s & \equiv &  \sum_{s_1,s_2} \partial
\cdot j_s \Big |_{s_1,s_2} + \sum_{s_1,s_2,s_3} \partial
\cdot j_s \Big |_{s_1,s_2, s_3} \\ & = &\sum_{s_1,s_2} \left(C_{s_1,s_2,s}(\tilde{\lambda}) \frac{1}{\tilde{N}} [j_{s_1}][j_{s_2}]\right)
+ \sum_{s_1,s_2,s_3}\left( C_{s_1,s_2,s_3,s}(\tilde{\lambda}) \frac{1}{\tilde{N}^2} [j_{s_1}][j_{s_2}][j_{s_3}]\right)\,, \label{anom}
\end{eqnarray}
where $[j_s]$ denotes $j_s$ or any of its conformal descendants, and $C_{s_1,s_2,s}$ and $C_{s_1,s_2,s_3,s}$
are numerical coefficients that depend on $s_1$, $s_2$ (and $s_3$) and also $\tilde{\lambda}$. The ``double-trace" operator $[j_{s_1}][j_{s_2}]$
appearing on the right-hand side can be fixed by conformal symmetry up to the overall normalization that can be absorbed in $C_{s_1,s_2,s}$,
as we will work out explicitly below. Similarly, one could fix the structure of the ``triple-trace" term. However, it is easy to see that
this term does not affect the anomalous dimension of $j_s$ to order $1/N$ or the planar 3-point functions, and we will ignore it below.

We can fix the $\tilde \lambda$-dependence of $C_{s_1,s_2,s}(\tilde{\lambda})$ by calculating
the correlation function of both sides of equation \eqref{anom} with $j_{s_1}$ and $j_{s_2}$. To leading order at large $N$,
the resulting correlator factorizes and we find
\begin{eqnarray}
	\langle j_{s_1} j_{s_2} \partial \cdot j_s \rangle & \sim &  \frac{1}{\tilde{N}} C_{s_1,s_2,s}(\tilde{\lambda}) \langle j_{s_1} j_{s_1} \rangle \langle j_{s_2} j_{s_2} \rangle\,.
\end{eqnarray}
On the other hand, from the results of \cite{MZ}, we have, see (\ref{CSbos-sss}), (\ref{CSbos-ss0}) and (\ref{CSfer-ss0}):
\begin{equation}
\begin{aligned}
&\langle j_{s_1} j_{s_2} \partial \cdot j_s \rangle \sim \tilde N \frac{\tilde\lambda}{1+\tilde\lambda^2}\,,\\
&\langle j_{s_1} j_{0} \partial \cdot j_s \rangle \sim \tilde N \tilde\lambda\,,\\
&\langle j_{s_1} \tilde{j}_{0} \partial \cdot j_s \rangle \sim \tilde N \tilde\lambda\,,
\end{aligned}
\end{equation}
where $\sim$ means equality up to $\tilde N$- and $\tilde \lambda$-independent numerical coefficients,
and this result follows from the fact that the current non-conservation can only arise from the parity violating terms in the 3-point functions
(\ref{CSbos-sss}), (\ref{CSbos-ss0}) and (\ref{CSfer-ss0}).
We also know that, see eq. (\ref{2pt-CSbos}) and (\ref{2pt-CSfer}): \footnote{Recall that our normalization of $j_0$ and $\tilde{j}_0$ differ from the one used in \cite{MZ}, where $\langle j_{0} j_{0}\rangle , \langle \tilde{j}_{0} \tilde{j}_{0}\rangle \sim \tilde N (1+\tilde{\lambda}^2)^{-1}$.}
\begin{equation}
\begin{aligned}
&\langle j_{s_1} j_{s_1}\rangle \sim \tilde N \,,\qquad s_1\neq 0\,, \\
&\langle j_{0} j_{0}\rangle \sim \tilde N (1+\tilde{\lambda}^2)\,,\qquad \langle \tilde{j}_{0} \tilde{j}_{0} \rangle\sim \tilde N (1+\tilde{\lambda}^2)\,. \label{normalization-j0}
\end{aligned}
\end{equation}
Putting everything together, we find
\begin{equation}
C_{s_1,s_2,s}(\tilde\lambda) \sim \frac{\tilde{\lambda}}{1+\tilde{\lambda}^2}= \left(\tilde{\lambda}+\frac{1}{\tilde{\lambda}}\right)^{-1}\,,
\label{C-of-lam}
\end{equation}
which is valid both for $s_1,s_2\neq 0$ and for the case when either one of $s_1$ or $s_2$ is zero (in the case $s_1=s_2=0$, we have $C_{0,0,s}=0$).
This $\tilde\lambda$-dependence holds both in the quasi-boson and quasi-fermion theories.

Via radial quantization (or equivalently directly using conformal invariance in flat space),
the form \eqref{anom} for the divergence of $j_s$ implies that the twist, $\tau_s=\Delta_s-s$ of $j_s$ takes the form, to
the leading order in $1/\tilde N$
\begin{eqnarray}
\!\!\!\tau_s-1 =\sum_{s_1 \neq 0} \left(C_{s_1,0,s}(\tilde{\lambda})\right)^2 \alpha_{s_1,0,s}
\frac{n_{s_1} n_{0}(1+\tilde\lambda^2)}{\tilde{N} n_s}+\sum_{s_1, s_2 \neq 0} \left(C_{s_1,s_2,s}(\tilde{\lambda})\right)^2 \alpha_{s_1,s_2,s}
\frac{n_{s_1} n_{s_2}}{\tilde{N} n_s}
\label{twistB}
\end{eqnarray}
in the quasi-boson theory, and a similar expression in the quasi-fermion theory, with $n_0$ replaced by $\tilde{n}_0$.
Here $\alpha_{s_1,s_2,s}$ and $\alpha_{s_1,0,s}$ are numerical coefficients that depend on the explicit
form of the ``double-trace" primaries on the right-hand side of \eqref{anom}, and $n_s, n_0, \tilde{n}_0$ are the 2-point normalization
coefficients defined in (\ref{js-norm-sc}), (\ref{j0-norm}) and (\ref{tj0-norm}). Note that the triple-trace component of the RHS of equation \eqref{anom} does not contribute to the anomalous dimension at the order $\frac{1}{\tilde{N}}$.

From \eqref{twistB}, we see that to order $1/\tilde N$ the twists take the form:
\begin{equation}
\label{twists-anom}
\tau_s -1 = \frac{1}{\tilde{N}} \left( a^{(F/B)}_s  \frac{\tilde{\lambda}^2}{1+\tilde{\lambda}^2} + b^{(F/B)}_s  \frac{\tilde{\lambda}^2}{(1+\tilde{\lambda}^2)^2} \right)\,,
\end{equation}
where the value of $a_s$ and $b_s$ depends on the spin $s$ only. A priori, the values of $a_s$ and $b_s$ may be different for the quasi-Fermionic theory and the quasi-Bosonic theory, hence the superscripts $F$ and $B$. Assuming the uniqueness of the parity violating terms in the 3-point functions of non-zero spin operators, one expects
from the analysis of \cite{MZ} that $b_s^{B}=b_s^{F}$. We will verify this explicitly from the calculations in section \ref{sec:classical}. Note that
the result $b_s^{B}=b_s^{F}$ is in fact necessary for the bose/fermi duality to work; this is because the calculation of $b_s$, or equivalently of $C_{s_1,s_2,s}$ with
$s_1,s_2\neq 0$, is identical in the regular CS-boson and critical CS-boson (the planar 3-point functions of non-zero spin operators are unaffected by the Legendre transform), and similarly in regular CS-fermion and critical CS-fermion. We will also find by our explicit calculations that $a_s^{B}=a_s^{F}$; this result appears to be 
more surprising, as it is not required by the bose/fermi duality.

\subsection{Constraining the divergence of $j_s$}
\label{constrain}

The divergence of $j_{s}$ must be a conformal primary to leading order in $1/N$. A straightforward argument for this is given in Appendix A of \cite{MZ}. Another simple way of seeing this is based on conformal representation theory \cite{Giombi:2011kc} -- at leading order in $1/N$, the primary operator $j_{s}$ has twist 1, and therefore heads a short representation $(\Delta,s)=(s+1,s)$ of the conformal group. When $1/N$ corrections are included, the primary $j_{s}$ acquires an anomalous dimension and now heads a long representation of the conformal group. To transform a short representation $(s+1,s)$ to a long representation, we require additional states, which must transform amongst themselves as a long representation $(s+2, s-1)$ to leading order in $1/N$. This long representation is headed by a primary operator, which is the divergence of $j_{s}$.

We denote the contribution of double-trace operators involving $j_{s_1}$ and $j_{s_2}$ to the RHS of \eqref{anom} by
\begin{equation}
   \partial \cdot j_s \Big|_{s_1,s_2}=C_{s_1,s_2,s} [j_{s_1}] [j_{s_2}]\,,
   \label{eq}
\end{equation}
where for convenience we have absorbed the factor of $1/\tilde N$ in (\ref{anom}) into $C_{s_1,s_2,s}$.
Below we explicitly determine the unique allowed combination of descendants of $j_{s_1}$ and $j_{s_2}$ represented by $[j_{s_1}][j_{s_2}]$ on the LHS of \eqref{eq} up to a single overall constant, $C_{s_1,s_2,s_3}$ by demanding that $\partial \cdot j_s \Big|_{s_1,s_2}$ is annihilated by the generator of special conformal transformations $K_\mu$ to leading order in $1/N$.

For simplicity, in this subsection we assume the null polarization vector $z^\mu$ always to be $\delta^\mu_-$, so $j_s(x,z) = (j_{s}){}_{---\ldots}=j_s^{+++\ldots}$. We also use $(j_{s}){}_\mu$ and $(j_{s}){}_{\mu\nu}$ to denote $(j_{s}){}_{\mu - - - \ldots}$ and $(j_{s}){}_{\mu\nu - - - \ldots}$ respectively.

\subsubsection{$s_1$ and $s_2$ nonzero} \label{sec1}
Let us first consider the case when both spins are nonzero: $s_i \neq 0$.

The scaling dimension of the LHS of Equation \eqref{eq} is $\Delta=s+2$. We match the scaling dimension in the RHS of \eqref{eq} by including $p=s+2-(s_1+1+s_2+1)=s-s_1-s_2$ derivatives in $[j_{s_1}][j_{s_2}]$. A general expression with $p$ derivatives acting on $j_{s_1}$ and $j_{s_2}$, is:
\begin{equation}
\sum_{n=0}^{p} c_n \partial^{\mu_1} \ldots \partial^{\mu_n} j_{s_1}^{\alpha_1 \ldots \alpha_{s_1}} \partial^{\nu_1} \ldots \partial^{\nu_{n-p}} j_{s_2}^{\beta_1 \ldots \beta_{s_2}} \label{eqa}\,.
 \end{equation}
Here, we wrote all free indices explicitly. This expression is symmetric with respect to permutations $\alpha_i \leftrightarrow \alpha_j$, $\beta_i \leftrightarrow \beta_j$,  $\mu_i \leftrightarrow \mu_j$, and $\nu_i \leftrightarrow \nu_j$.

We must now contract each term in expression \eqref{eqa} with dimensionless tensors. These can come from the following lists:
\begin{eqnarray}
	\text{List 1:}&~& \eta_{-\alpha_i}, ~ \eta_{-\beta_i}, ~ \eta_{-\mu_i}, ~ \eta_{-\nu_i} \\
	\text{List 2:}&~& \epsilon_{\alpha_i \beta_j -}, ~ \epsilon_{\alpha_i \mu_j -}, ~\epsilon_{\mu_i \beta_j -}, ~ \epsilon_{\alpha_i \nu_j -}, \epsilon_{\nu_i \beta_j -}, ~ \epsilon_{\mu_i \nu_j -} \\
	\text{List 3:}&~& \eta^{\mu_i \mu_j}, ~\eta^{\nu_i \nu_j}, \ldots \\
	\text{List 4:}&~& \epsilon^{\alpha_i \beta_j \mu_k}, ~ \epsilon^{\alpha_i \beta_j \nu_k}, \ldots
\end{eqnarray}
Let us contract equation \eqref{a} with $n_1$ tensors from List 1, $n_2$ tensors from List 2, $n_3$ tensors from List 3 and $n_4$ tensors from list 4.

Because the total spin of $[j_{s_1}][j_{s_2}]$ must be $s-1$,
we require $n_1+n_2=s-1$. (Recall that we take all free indices in $\partial \cdot j_s$ to be in the $-$ direction, so the spin is simply the number of lower $-$ indices.)
The total number of free indices in \eqref{eqa} is $p+s_1+s_2=s$; each tensor from List 1 removes one free index, each tensor from List 2 or 3
removes two free indices, and each tensor from List 4 removes 3 indices, so we also require $n_1+2n_2+2n_3+3n_4=s$. This implies $n_2+2n_3+3n_4=1$, which then implies $n_2=1$, $n_1=s-2$ (and $n_3=n_4=0$).

Hence we require $s-2$ tensors from List 1 and $1$ tensor from List 2. Choosing the tensor from List 2 automatically fixes which tensors from List 1 we need to use. Note that the resulting operators always involve the $\epsilon$-tensor, illustrating the fact that the breaking of current conservation in 3-point functions arises from parity violating terms.

Contracting each of the six tensors in List 2 with equation \eqref{eqa} yields:
\begin{equation}
\begin{split}
      \partial \cdot j_s \Big|_{s_1,s_2} & =   \sum_{n=0}^p \epsilon_{\mu\nu-} \Big(a_n \partial_-^n j^\mu_{s_1} \partial_-^{p-n} j^\nu_{s_2}  + b_n \partial_-^{n-1} \partial^\nu j^\mu_{s_1}\partial_-^{p-n} j_{s_2} +
  \\  &  +c_n \partial_-^{n-1} \partial^\mu j_{s_1}\partial_-^{p-n} j^\nu_{s_2} + d_n \partial_-^{n} j^\mu_{s_1} \partial_-^{p-n-1} \partial^\nu j_{s_2}
  \\  & + e_n \partial_-^n j_{s_1} \partial_\mu \partial_-^{p-n-1} j_{s_2}^\nu   + f_n \partial_-^{n-1}\partial^\mu j_{s_1}\partial_-^{p-n-1}\partial^\nu j_{s_2} \Big)\,.
      \label{gen-nonconservationa}
\end{split}
\end{equation}
However the six types of terms in \eqref{gen-nonconservationa} are not linearly independent, as one can check by explicitly writing out the sums over $\mu$ and $\nu$. We can choose a basis of three linearly independent terms and write the most general form for $[j_{s_1}][j_{s_2}]$ with correct scaling dimension and spin as:
\begin{equation}
    	\partial \cdot j_s \Big|_{s_1,s_2} = \sum_{n=0}^p \epsilon_{\mu\nu-}\left(a_n \partial_-^n j^\mu_{s_1} \partial_-^{p-n} j^\nu_{s_2} + b_n \partial_-^{n-1} \partial^\nu j^\mu_{s_1}\partial_-^{p-n} j_{s_2} + e_n \partial_-^n j_{s_1} \partial^\mu \partial_-^{p-n-1} j_{s_2}^\nu \right)\,, \label{gen-nonconservation}
\end{equation}
where $b_0=0$ and $e_p=0$. We also must require that, when we interchange the spins  $s_1 \leftrightarrow s_2$, $b_n \leftrightarrow -e_{p-n}$ and $a_n \leftrightarrow -a_{p-n}$ .

Next we apply the constraint that the expression be a conformal primary. Acting on this expression with $K_3$ and $K_+$, as illustrated in Appendix \ref{appendix-constrain}, we are able to determine $a_n$, $b_n$ and $e_n$ up to one undetermined constant $C_{s_1,s_2,s}$.
\begin{eqnarray}
a_n & = & C_{s_1,s_2,s} \frac{(-1)^{n+1} \left(s_1
   (n-s+s_1-s_2)+s_2 (n+2 s_1)
   (-1)^{s+s_1+s_2}\right)
   }{(s-s_1-s_2)
   (s+s_1+s_2)} \binom{s-s_1-s_2}{n}
   \binom{s+s_1+s_2}{n+2
   s_1} \nonumber\,, \\
b_{n} & = & C_{s_1,s_2,s} (-1)^n \binom{s-s_1-s_2-1}{n-1}\binom{s+s_1+s_2}{n+2s_1} \nonumber\,, \\
e_n & = & C_{s_1,s_2,s} (-1)^{s-s_1-s_2+n+1} \binom{s-s_1-s_2-1}{n} \binom{s+s_1+s_2}{n+2s_1}\,. \label{gendiv}
\end{eqnarray}
This formula is also valid if $s_1=s_2$.

\subsubsection{$s_2=0$, Quasi-Fermionic}
Let us next consider the contribution to the non-conservation equation from $j_{s_1}$ and $\tilde{j}_{0}$ in the quasi-fermionic theory.

In this case, $[j_{s_1}][\tilde{j}_0]$ requires $p=s-s_1-1$ derivatives, $s-1$ tensors from List 1, and no tensors from the other lists. Hence, we have the following expression:
\begin{equation}
	 \partial \cdot j_s \Big|_{s_1,0}=\sum_{m=0}^p c_m \partial_{-}^m j_{s_1} \partial_-^{p-m} \tilde{j}_{0} \,.
	 \label{eq2}
\end{equation}

Now we apply the constraint that the expression be a conformal primary, as illustrated in appendix \ref{appendix-constrain}. We find that
\begin{equation}
 c_m = \frac{-(m-p-1) (m-p-2)}{   m (m+ 2 s_1)}c_{m - 1},
\end{equation}
which can be solved to give
\begin{equation}
    c_m  = \left( -1\right)^m \binom{s-s_1}{m} \binom{s+s_1-1}{m+2s_1}  C_{s_1,\tilde{0},s}. \label{qfdiv}
\end{equation}
One can check that this form agrees with the various divergences calculated explicitly in \cite{Giombi:2011kc}, as well as those calculated in \cite{Skvortsov:2015pea,Giombi:2016hkj}.

\subsubsection{$s_2=0$, Quasi-Bosonic}
We now consider the contribution from $j_{s_1}$ with $s_1 \neq 0$ and $j_0$ to the divergence of $j_s$ in the quasi-bosonic theory, where the scalar primary has scaling dimension $1$. Here, the analysis of section \ref{sec1} applies again, but there are only three relevant tensors in List 2, of which only two are independent, yielding:
\begin{equation}
    	\partial \cdot j_s \Big|_{s_1,s_2} = \sum_{n=0}^p \epsilon_{\mu\nu-}\left(b_n \partial_-^{n-1} \partial^\nu j^\mu_{s_1}\partial_-^{p-n} j_{0} + f_n \partial_-^n j^\mu_{s_1} \partial^\nu \partial_-^{p-n-1} j_{0} \right)\,, \label{qb-nonconservation}
\end{equation}
where $p=s-s_1$.

Requiring that the expression is annihilated by $K_\delta$ gives:
\begin{eqnarray}
b_n & = & C_{s_1,0,s} (-1)^{n} \binom{s+s_1}{n+2s_1}\binom{s-s_1-1}{n-1}\,, \label{qbdiv1}\\
f_n & = & C_{s_1,0,s} (-1)^{n}\frac{s_1}{s+s_1} \binom{s+s_1}{n+2s_1}\binom{s-s_1-1}{n} \,,\label{qbdiv2}
 \end{eqnarray}
with $f_p=0$ and $b_0=0$. We checked that this matches the divergence of $j_4$ calculated explicitly in \cite{Aharony:2011jz}.

\subsection{The anomalous dimensions}
We can now use the explicit form of the non-conservation equation to determine the anomalous dimensions of the higher-spin operators
to order $1/N$. Using the index-free notation in terms of the null polarization vector $z$, we can write the non-conservation equation as
\be
\label{descgen}
\partial_{\mu} D^{\mu}_z j_s(x,z)={\cal K}_{s-1}(x,z)\,,
\ee
where $D^{\mu}_z$ is the operator defined in (\ref{Dz}). Recall that the two-point function a spin $s$ primary operator
of dimension $\Delta_s$ is
fixed by conformal invariance to be
\be
\label{conftwopoint}
\langle j_s (x_1,z_1) j_{s}(x_2,z_2)\rangle =
{\cal N}_s \frac{\left(\frac{z_1\cdot x_{12} z_2\cdot x_{12}}{x_{12}^2}-\frac{1}{2}z_1\cdot z_2\right)^{s}}{(x_{12}^2)^{\Delta_{s}}}\,,
\ee
where $z_1$, $z_2$ are two polarization vectors. Writing $\Delta_s = s+1+\gamma_s$
and taking the divergence on $x_1$ and $x_2$ on both sides of this equation,
one may derive the following formula for the anomalous dimension, valid to leading order in the breaking parameter \cite{Anselmi:1998ms, Belitsky:2007jp}
\begin{equation}
\label{mainform}
\gamma_s = -\frac{1}{s^2(s^2-\frac{1}{4})}
 \frac{(z\cdot x)^2\langle {\cal K}_{s-1} (x,z){\cal K}_{s-1}(0,z)\rangle_0}{\langle j_s (x,z) j_s(0,z)\rangle_0}\,,
\end{equation}
where the subscript `0' means that the correlators are computed in the ``unbroken" theory (in our case, to leading order at large $N$).

Let us define
\begin{equation}
\begin{aligned}
&{\cal K}_{s-1}^{(a)} = \sum_{s_1} C_{s_1,0,s} [j_{s_1}][j_0]\,,\qquad
\tilde{\cal K}_{s-1}^{(a)} = \sum_{s_1} C_{s_1,\tilde{0},s} [j_{s_1}][\tilde{j}_0]\,,\\
&{\cal K}_{s-1}^{(b)} =\sum_{s_1,s_2\neq 0} C_{s_1,s_2,s} [j_{s_1}][j_{s_2}]\,,
\end{aligned}
\end{equation}
so that in the quasi-boson theory we have $\partial \cdot j_s = {\cal K}_{s-1}^{(a)} +{\cal K}_{s-1}^{(b)}$, and in the
quasi-fermion $\partial \cdot j_s = \tilde{{\cal K}}_{s-1}^{(a)}+{\cal K}_{s-1}^{(b)}$. Using the explicit form of
these double-trace operators given in (\ref{qb-nonconservation}), (\ref{eq2}) and (\ref{gen-nonconservation}), and computing their two-point functions
using (\ref{2pt-CSbos}) and (\ref{2pt-CSfer}),\footnote{To compute the two-point functions of currents with one ``open" index,
one may take derivatives of (\ref{conftwopoint}) with respect to the polarization vectors.} we find
\begin{eqnarray}
&&\frac{(z\cdot x)^2\langle {\cal K}_{s-1}^{(a)}{\cal K}_{s-1}^{(a)}\rangle_0}{\langle j_s j_s \rangle_0} =
\tilde N (1+\tilde{\lambda}^2) \sum_{s_1} \frac{s^3((s-1)!)^2}{128\pi^2(s^2-s_1^2)(s_1!)^2} (C_{s_1,0,s})^2\,,\cr
&&\frac{(z\cdot x)^2\langle \tilde{\cal K}_{s-1}^{(a)}\tilde{\cal K}_{s-1}^{(a)}\rangle_0}{\langle j_s j_s \rangle_0} =
-\tilde N (1+\tilde{\lambda}^2) \sum_{s_1} \frac{(s!)^2 (s-s_1)}{128 \pi^2 s (s_1!)^2 (s+s_1)} (C_{s_1,\tilde 0,s})^2
\label{KKnorm}\,, \\
&&\frac{(z\cdot x)^2\langle {\cal K}_{s-1}^{(b)}{\cal K}_{s-1}^{(b)}\rangle_0}{\langle j_s j_s \rangle_0} =
\tilde N \sum_{s_1,s_2 \neq 0} \frac{s^3 ((s-1)!)^2 (s-s_1-s_2-1)! (s+s_1+s_2-1)!}{64\pi ^2 (s_1!)^2 (s_2!)^2 (s+s_1-s_2)! (s-s_1+s_2)!}
(C_{s_1,s_2,s})^2\,.\nonumber
\end{eqnarray}
A direct calculation using the equations of motion, described in the next section, and the result (\ref{C-of-lam}), allow us to fix the
undetermined ``structure constants" to be
\begin{equation}
C_{s_1,s_2,s}^B = -C_{s_1,s_2,s}^F = \frac{1}{\tilde N} \frac{\tilde\lambda}{1+\tilde\lambda^2}\cdot
\begin{cases}
\frac{32 i (s+s_1-s_2)!(s-s_1+s_2)!s_1!s_2!}{(s+s_1+s_2-1)!(s-s_1-s_2-1)!s!} & s_1+s_2=s-2,s-4,\ldots\,,\\
0 & {\rm otherwise}
\end{cases}
\label{Csss}
\end{equation}
\begin{equation}
C_{s_1,0,s} = \frac{1}{\tilde N} \frac{\tilde\lambda}{1+\tilde\lambda^2} \cdot
\begin{cases}
\frac{32i(s^2-s_1^2)s_1!}{s!} & s_1 =s-2,s-4,\ldots\,, \\
0 & {\rm otherwise}
\end{cases}
\label{Css0}
\end{equation}
\begin{equation}
C_{s_1,\tilde{0},s} = \frac{1}{\tilde N} \frac{\tilde\lambda}{1+\tilde\lambda^2} \cdot
\begin{cases}
\frac{32(s+s_1) s_1 !}{(s-1)!} & s_1 =s-2,s-4,\ldots \,, \\
0 & {\rm otherwise}
\end{cases}
\label{tCss0}
\end{equation}
Plugging these into (\ref{KKnorm}) and using the formula (\ref{mainform}), we find that the anomalous dimensions take the form (\ref{twists-anom}), with
$a_s^B=a_s^F=a_s$ and $b_s^B=b_s^F=b_s$ given by
\begin{equation}
a_s = \sum_{s_1 = s-2,s-4,\ldots } \frac{32 (s^2-s_1^2)}{\pi ^2 s \left(4 s^2-1\right)} =
\begin{cases}
\frac{16 (s-2)}{3 \pi ^2 (2 s-1)}\,, & s ~~  \rm{even}\,,\\
\frac{32 (s^2-1)}{3 \pi ^2 (4s^2-1)}\,,& s ~~ \rm{odd}\,,
\end{cases}
\label{as}
\end{equation}
and
\begin{equation}
\begin{aligned}
b_s &= \sum_{s_1+s_2 = s-2,s-4,\ldots}\frac{64(s+s_1-s_2)! (s-s_1+s_2)!}{\pi ^2 s\left(4 s^2-1\right) (s-s_1-s_2-1)! (s+s_1+s_2-1)!}\\
&=
\begin{cases}
\frac{2}{3\pi^2}\left(3\sum_{n=1}^s \frac{1}{n-1/2}+\frac{-38s^4+24s^3+34 s^2-24 s-32}{(4s^2-1)(s^2-1)}\right)\,,& s ~~ \rm{even}\,,\\
\frac{2}{3\pi^2}\left(3\sum_{n=1}^s \frac{1}{n-1/2}+\frac{-38s^2+20}{4s^2-1}\right)\,,& s ~~  \rm{odd}\,.
\end{cases}
\end{aligned}
\label{bs}
\end{equation}

Let us note here that it is straightforward to adapt the above results to the case of $O(N)$ gauge group: one simply drops all the odd-spins from
the sum. Doing this, we find
\begin{equation}
\begin{aligned}
&a_s^{O(N)} = \frac{16 (s-2)}{3 \pi ^2 (2 s-1)}\,,\\
&b_s^{O(N)} = \frac{4}{\pi ^2}{ \left(-2 \sum _{n=1}^{\frac{s}{2}-1} \frac{1}{n-\frac{1}{2}}+\frac{9}{4} \sum _{n=1}^{s-1} \frac{1}{n-\frac{1}{2}}
-\frac{59 s^4+18 s^3-4 s^2+54 s+35}{6 (4s^2-1)(s^2-1)}\right)}\,,
\label{as-bs-ON}
\end{aligned}
\end{equation}
and the anomalous dimensions take the same form as in (\ref{twists-anom}), with $\tilde N^{O(N)} = N(1+O(\lambda^2))$
and $\tilde \lambda^{O(N)} = \frac{\pi}{2}\lambda+O(\lambda^3)$. Note that $b_s^{O(N)}$ vanishes for $s=2$ and also for $s=4$, because
in the $O(N)$ case the divergence of $j_4$ can only take the form (\ref{dj4-MZ}).

\section{Current non-conservation from classical equations of motion} \label{sec:classical}
\newcommand{\pl}{\partial}
\subsection{CS-boson}
The generating function $J_{\rm b}(x,z)$ of the higher-spin operators in the CS-boson theory was given in (\ref{Jb-CS}). Since we will be working
to leading order in $1/k$, it will be sufficient to expand the generating function to linear order in the gauge field. We note that
\begin{equation}
\begin{aligned}
(\deh - i \hat{A})^n\phi  & = \deh^n \phi- i \sum_{k=0}^{n-1} \deh^k \hat{A} \deh^{n-1-k}\phi + O(A^2) \\
&=  \deh^n\phi - i \sum_{k=0}^{n-1} (\deh_x+\deh_y)^k  \deh^{n-1-k}_x \hat{A}(y)\phi(x)|_{y\rightarrow x} + O(A^2) \\
& = \deh^n\phi - i\frac{(\deh_x+\deh_y)^n - \deh_x^n}{\deh_y} \hat{A}(y)\phi(x)|_{y\rightarrow x}+ O(A^2)\,.
\end{aligned}
\end{equation}
Since this expression involves the same power $n$ everywhere, we can extend this formula to any function of $\hat D$ acting on $\phi$:
\begin{equation}
F(\deh - i \hat{A})\phi = F(\deh)\phi - i \frac{F(\deh_x+\deh_y)-F(\deh_x)}{\deh_y}\hat{A}(y)\phi(x)|_{y\rightarrow x}+O(A^2)\,.
\end{equation}
A similar result applies when a power of the covariant derivative acts on $\bar\phi$.  Therefore, to linear order in the gauge field,
the generating function of the higher-spin operators is
\begin{equation}
\begin{aligned}
&J_{\rm b}= f_{\rm b}(\deh_1,\deh_2)\bar{\phi}(x_1)\phi(x_2)+ig(\deh_1,\deh_2,\deh_3)\bar{\phi}(x_1)\hat A(x_3)\phi(x_2)\,,\\
&g(u,v,w)=\frac{f_{\rm b}(u+w,v)-f_{\rm b}(u,v+w)}{w}\,,\qquad f_{\rm b}(u,v)=e^{u-v}\cos(2\sqrt{uv})\,,
\end{aligned}
\end{equation}
where in the first line it is understood that after taking the derivatives all points are set to $x$.

To calculate the divergence of the spin $s$ operators, we should evaluate
\begin{equation}
\partial_{\mu}D^{\mu}_z J_{\rm b}(x,z) \equiv \partial \cdot J_{\rm b}\,.
\end{equation}
When the operator $\partial_{\mu}D^{\mu}_z$ acts on the $A$-independent piece of $J_{\rm b}$, one gets \cite{Skvortsov:2015pea,Giombi:2016hkj}
\begin{equation}
\begin{aligned}
&\partial_{\mu}D^{\mu}_z f_{\rm b}(\deh_1,\deh_2)\bar{\phi}(x_1)\phi(x_2)
= \left[h(\deh_1,\deh_2)\partial_1^2+\tilde h(\deh_1,\deh_2)\partial_2^2\right]\bar{\phi}(x_1)\phi(x_2)\,,\\
&h(u,v) = (\frac{1}{2} \pl_u + \frac{u-v}{2} \pl^2_u + v\pl_{uv}) f(u,v)\,,\\
&\tilde{h} (u,v) =(\frac{1}{2} \pl_v+ \frac{v-u}{2} \pl^2_v + u\pl_{uv}) f(u,v)\,.
\label{dj-A0}
\end{aligned}
\end{equation}
In the interacting theory, the equations of motion to linear order in $A$ are
\begin{equation}
\begin{aligned}
&\partial^2 \phi= i(\pl\cdot A)\phi +2i A\cdot \pl \phi\,,\\
&\partial^2 \bar{\phi}= -i\bar{\phi}(\pl\cdot A) -2 i (\pl \bar{\phi})\cdot A\,.
\end{aligned}
\end{equation}
Using this into (\ref{dj-A0}), and combining with the term that arises when $\partial^{\mu}D^{\mu}_z$ acts on the piece of $J_{\rm b}$ linear in $A$ (where
one can just use the free equation of motion $\partial^2\phi=0$),
one should find a gauge invariant result. We have explicitly checked that all the terms involving $\hat{A}$ indeed cancel out, and one is left with
terms involving only the field strength $F=dA$. The final result takes the form
\begin{equation}
\pl \cdot J_{\rm b}=\left[k_1(\hat \partial_1,\hat\partial_2,\hat\partial_3) \pl_1^\mu+k_2(\hat \partial_1,\hat\partial_2,\hat\partial_3) \pl_2^\mu+k_3(\hat \partial_1,\hat\partial_2,\hat\partial_3) \pl_3^\mu\right] \bar{\phi}(x_1) (iF_{\mu\rho}(x_3)z^\rho) \phi(x_2)\,,
\label{dJb}
\end{equation}
where
\begin{align}
k_1(u,v,w)&=\frac{2}{w}h(u+w,v)- \frac{1}{w} \big(\frac{1}{2} - (v+w) \pl_u + v \pl_v + w\pl_w\big)g(u,v,w)\notag\,,\\
k_2(u,v,w)&=-\frac{2}{w}\tilde{h}(u,v+w)-\frac{1}{w}\big(\frac{1}{2} + u \pl_u - (u+w) \pl_v + w \pl_w\big)g(u,v,w)\,,\\
k_3(u,v,w)&=\frac{1}{w}(h(u+w,v)-\tilde{h}(u,v+w)) -\frac{1}{w}\big(\frac{1}{2} +u\pl_u +v\pl_v -(u+v)\pl_w\big)g(u,v,w)\,.\notag\end{align}

We can now use the equation of motion for $A_\mu$, which reads (to linear order in $A$)
\begin{equation}
 \frac{k}{4\pi} \epsilon^{\mu\nu\rho} (F_{\nu\rho})^i_{\ j} = (\partial^{\mu}\bar\phi_j)\phi^i-\bar\phi_j\partial^{\mu}\phi^i\,,
\end{equation}
or equivalently
\begin{equation}
(F_{\mu\rho})^{i}_{\ j}= \frac{2\pi}{k} \epsilon_{\mu\rho\nu} \left( (\partial^{\nu}\bar\phi_j)\phi^i-\bar\phi_j\partial^{\nu}\phi^i\right)\,.
\end{equation}
After plugging this into (\ref{dJb}), we get
\begin{align}
\pl \cdot J_{\rm b}&=-\frac{2\pi i}{k} \epsilon_{\mu\nu\rho} z^\rho \left[k_1 \pl_1^\mu+k_2 \pl_2^\mu+k_3 (\pl_3^\mu + \pl_4^\mu) \right] (\pl_3^\nu-\pl_4^\nu )\bar{\phi}_i(x_1) \phi^i (x_4) \bar{\phi}_j (x_3) \phi^j(x_2) \,.
\label{dJbfinal}
\end{align}
Note that we had to point-split $\pl_3\rightarrow \pl_3 +\pl_4$. To make contact with the analysis of the previous section, one should express
this as a sum of double-trace primaries. Note that the scalar bilocals with derivatives acting on them can be expressed as
linear combinations of the higher-spin operators and their derivatives. Doing this, one finds precisely the decomposition derived
in the previous section
\begin{equation}
\partial \cdot J_{\rm b} ={\cal K}^{(a)}+{\cal K}^{(b)}\,,
\label{dj-regbos}
\end{equation}
where ${\cal K}^{(a)}\sim \sum_{s_1} C_{s_1,0,s}[j_{s_1}][j_0]$ and ${\cal K}^{(b)}\sim \sum_{s_1,s_2}C_{s_1,s_2,s}[j_{s_1}][j_{s_2}]$
are the double-trace operators given respectively in (\ref{qb-nonconservation}) and (\ref{gen-nonconservation}),
and the $C_{s_1,s_2,s}$ coefficients are determined to be
\begin{equation}
C_{s_1,s_2,s} = +\frac{2\pi i\lambda}{N}  \frac{4(s+s_1-s_2)!(s-s_2+s_1)!}{(s+s_1+s_2-1)!(s-s_1-s_2-1)!}\frac{s_1!s_2!}{s!}\,, \qquad s_1+s_2=s-2j\,,~j>0\,,
\label{Cb-regbos}
\end{equation}
and
\begin{equation}
C_{s_1,0,s} = 4\frac{2\pi i\lambda}{N}  \frac{(s^2-s_1^2)s_1!}{s!}\,,\qquad s_1=s-2j\,,~j>0\,.
\label{Ca-regbos}
\end{equation}
Extending these to all orders in $\lambda$ by sending $\lambda\rightarrow \frac{2}{\pi} \tilde\lambda/(1+\tilde\lambda^2)$ and $N\rightarrow \tilde N/2$, one obtains
the results quoted in (\ref{Csss}) and (\ref{Css0}).

Note that from the form (\ref{dJbfinal}) of the divergence, it is also
straightforward to compute the anomalous dimensions by directly using Wick contractions of $\phi$ and the master formula (\ref{mainform}). In fact,
this allows to obtain the anomalous dimensions to order $\lambda^2$ and exactly in $N$. We find for $s=1,2,3,\ldots$:
\begin{equation}
\gamma_s = \frac{\lambda^2}{N}\left\{0,0,\frac{32}{105}+\frac{8}{105 N},\frac{12}{35}+\frac{4}{105N},
\frac{1504}{3465}+\frac{24}{385 N},\frac{4192}{9009}+\frac{32}{693 N},\ldots\right\}+O(\lambda^3)\,.
\label{gams-sc}
\end{equation}
This takes the form
\begin{equation}
\gamma_s = \frac{\pi^2\lambda^2}{2N}(a_s+b_s)+\frac{\lambda^2}{N^2}\gamma_s^{(2)}+O(\lambda^3)\,,
\end{equation}
where $a_s$ and $b_s$ are given in (\ref{as}) and (\ref{bs}), and the coefficients $\gamma_s^{(2)}$ at order $\lambda^2/N^2$ can be found to be:
\begin{align}\label{subleadB}
\begin{aligned}
\gamma^{(2)}_s
&=
\begin{cases}
2 \left(H_{s-\frac{5}{2}}-H_{\frac{s-3}{2}}\right)-\frac{4 (s-2) \left(4 s^2-4 s+3\right)}{3 (s-1) (2 s-3) (2 s-1)}\,,& s ~~ \rm{even}\,,\\
2 \left(H_{s-\frac{3}{2}}-H_{\frac{s}{2}-1}\right)-\frac{2 (s-1) \left(8 s^2+8 s+3\right)}{3 s (2 s-1) (2 s+1)}\,,& s ~~  \rm{odd}\,,
\end{cases}
\end{aligned}
\end{align}
where $H_n$ is the harmonic number. We note that the dimensions of even spin currents differ by a simple fraction from that of the odd spin ones. We also observe that, unlike the
order $1/N$ term, these coefficients do not display logarithmic behavior at large spin.

\subsubsection{Critical boson}
Let us now study the critical boson theory obtained by adding the $(\bar\phi\phi)^2$ interaction and flowing to the IR.
As reviewed earlier, the $1/N$ expansion of the CFT can be developed using the action (\ref{crit-bos}). In the IR, $\sigma_b$ becomes a
scalar primary with $\Delta=2+O(1/N)$, and the $\sigma_b$ equation of motion formally removes $\bar\phi\phi$ from the spectrum.

It is evident that the equations of motion and hence the divergence of the higher-spin currents
will be modified due to the interaction with $\sigma_b$ (the form of the currents themselves stay the same as in (\ref{Jb-CS})).
Working to linear order in the gauge field, the equations of motion are modified to
\begin{equation}
\begin{aligned}
&\partial^2 \phi= i(\pl\cdot A)\phi +2i A\cdot \pl \phi+\frac{1}{N}\sigma_b\phi\,, \\
&\partial^2 \bar{\phi}= -i\bar{\phi}(\pl\cdot A) -2 i (\pl \bar{\phi})\cdot A+\frac{1}{N}\sigma_b\bar\phi\,.
\end{aligned}
\end{equation}
Consequently, when computing the divergence of $J_{\rm b}$, the descendant acquires an additional term linear in $\sigma$, and to
leading order in $1/N$ and $1/k$ is given by
\begin{equation}
\begin{aligned}
&\partial \cdot J_{\rm b} = {\cal K}_{\rm reg.CS-bos.}+{\cal K}_{\rm crit.bos.}\,,\\
&{\cal K}_{\rm crit.bos.} = \frac{1}{N}(h(\hat\partial_1+\hat\partial_3,\hat\partial_2)
+ \tilde{h}(\hat\partial_1,\hat\partial_2+\hat\partial_3)) \bar{\phi}(x_1) \phi(x_2) \sigma_b(x_3)\,,
\end{aligned}
\end{equation}
where $h(u,v)$ and $\tilde h(u,v)$ were defined in (\ref{dj-A0}), and ${\cal K}_{\rm reg.CS-bos.}$ is the descendent computed in the previous section, given
in (\ref{dJb}). To get the final result for the divergence, one should still impose that $\bar\phi\phi=0$ as a consequence of the equation of motion
for $\sigma_b$. This means that we should drop the term ${\cal K}^{(a)} \sim C_{s_1,0,s}\sum_{s_1} [j_0][j_{s_1}]$ from ${\cal K}_{\rm reg.CS-bos.}$.
Also writing ${\cal K}_{\rm crit.bos.}$ in terms of primaries and dropping all the
$\bar\phi\phi$ terms, one finds the final result
\begin{equation}
\begin{aligned}
&\partial \cdot J_{\rm b} = \tilde{\cal K}^{(a)}+{\cal K}^{(b)}\,,\\
&\tilde{\cal K}^{(a)} =
\sum_{s_1=s-2,s-4,\ldots} C_{s_1,\tilde{0},s} \sum_{m=0}^{s-s_1-1} \left( -1\right)^m \binom{s-s_1}{m} \binom{s+s_1-1}{m+2s_1}
 \hat\partial^m j_{s_1} \hat \partial^{s-s_1-1-m} \sigma_b\,, \\
& C_{s_1,\tilde{0},s} =\frac{ 2(s+s_1)s_1!}{(s-1)!}\frac{1}{N}\,.
\end{aligned}
\end{equation}
where ${\cal K}^{(b)}\sim \sum_{s_1,s_2}C_{s_1,s_2,s}[j_{s_1}][j_{s_2}]$ remains the same as in the regular CS-boson theory of the previous section.
Note that $\tilde{\cal K}^{(a)}$ has precisely the form predicted by conformal symmetry for the quasi-fermion theory, eq. (\ref{eq2}). Defining
$\sigma_b = 4\pi\lambda \tilde j_{0}^{\rm crit.bos.}$, this result can be seen to be precisely related by the bose/fermi duality to the
divergence in the CS-fermion theory, which we compute in the next section.

Extending the above result to all orders in $\tilde\lambda$ by using the arguments in section \ref{general_analysis}, one can deduce that
the anomalous dimensions in the critical boson model coupled to Chern-Simons are
\begin{equation}
\label{gams-crit}
\gamma_s^{\rm crit.} = \frac{1}{\tilde N} \left(\frac{1}{1+\tilde\lambda^2}a_s+\frac{\tilde\lambda^2}{(1+\tilde\lambda^2)^2}b_s \right)\,.
\end{equation}

\subsection{CS-fermion}
\label{sec:CSF}
The generating function of the higher-spin operators in the CS-fermion theory was given in (\ref{Jf-CS}). Linearizing it in the gauge field,
as described in the boson case above, we find
\begin{equation}
\begin{aligned}
&J_{\rm f}= f_{\rm f}(\deh_1,\deh_2)\bar{\psi}(x_1)\hat{\gamma}\psi(x_2)+ig(\deh_1,\deh_2,\deh_3)\bar{\psi}(x_1)\hat{\gamma}\hat{A}(x_3)\psi(x_2)\,,\\
&g(u,v,w)=\frac{f_{\rm f}(u+w,v)-f_{\rm f}(u,v+w)}{w}\,,\qquad f_{\rm f}(u,v)= \frac{e^{u-v} \sin \left(2 \sqrt{u v} \right)}{2\sqrt{u v} }\,.
\label{Jf-lin}
\end{aligned}
\end{equation}
The equations of motion to linear order in the gauge field are
\begin{equation}
\begin{aligned}
&\slashed{\pl}\psi = i \slashed{A} \psi \,,\qquad \pl_{\mu}\bar\psi \gamma^{\mu} = -i\bar\psi \slashed{A}\,,\\
&\partial^2 \psi=\frac{i}{2} \gamma^{\mu\nu}  F_{\mu\nu}\psi +i (\pl\cdot A)\psi+2i A\cdot \pl \psi\,,\\
&\partial^2 \bar{\psi}=\frac{i}{2} \bar{\psi}\gamma^{\mu\nu} F_{\mu\nu} -i\bar{\psi}\pl \cdot A-2 i (\pl^{\mu} \bar{\psi})A_{\mu}\,.
\end{aligned}
\end{equation}

We are now prepared to evaluate the divergence $\partial \cdot J_{\rm f}$. The calculation will consist of two terms essentially. The first one arises from acting with $\pl_{\mu} D_z^{\mu}$ on the $A$-independent part of (\ref{Jf-lin}), and which gives terms proportional to the ``descendant operators" $\pl^{\mu}\bar{\psi}\gamma_{\mu}$, $\slashed{\pl} \psi$ and $\partial^2{\bar{\psi}}$, $\partial^2{\psi}$, which are non-zero in the interacting fermion theory:
\begin{eqnarray}
&& \pl_{\mu} D_z^{\mu} f_{\rm f}(\deh_1,\deh_2)\bar{\psi}(x_1)\hat{\gamma}\psi(x_2)
= \left[\slashed{\pl}_1 q(\deh_1,\deh_2) + \slashed{\pl}_2  \tilde{q}(\deh_1,\deh_2) +\hat{\gamma} \partial^2_1   h(\deh_1,\deh_2)
+ \hat{\gamma} \partial^2_2  \tilde{h}(\deh_1,\deh_2) \right]\bar{\psi}(x_1)\psi(x_2)\,,  \cr
&& q(u,v)=\big( \frac{1}{2} f_{\rm f}+ v (\partial_v f_{\rm f} - \partial_u f_{\rm f}) \big) \,,\qquad
\tilde q(u,v)= \big( \frac{1}{2} f_{\rm f} + u (\partial_u f_{\rm f} - \partial_v f_{\rm f}) \big)\,,\cr
&&h(u,v)= \big(\frac{3}{2} \partial_u f_{\rm f} + \frac{u-v}{2} \partial^2_u f_{\rm f} + v \partial_{u v} f_{\rm f} \big)\,,\\
&&\tilde h(u,v)=\big(\frac{3}{2} \partial_v f_{\rm f} + \frac{v-u}{2} \partial^2_v f_{\rm f} + u \partial_{u v} f_{\rm f} \big)\nonumber
\label{nonconsgen}
\end{eqnarray}

The second term is the result of acting with $\pl_{\mu} D_{z}^{\mu}$ on the piece of (\ref{Jf-lin}) proportional to $A$ (in this piece, we can use the free Dirac
equation of motion). To simplify the calculation, one may impose the $\hat{A}=0$ ``light-cone" gauge after differentiation with respect to the $z_{\mu}$ is carried out everywhere. The full form of the descendant as a function of $F_{\mu\nu}$ can be then reconstructed using gauge invariance. As a consistency check, we have also performed
the calculation in arbitrary gauge, and verified that all unwanted $\hat{A}$ terms drop out. The final result takes the form
\begin{equation}\notag
\begin{aligned}
&\pl \cdot J_{\rm f}=\left[k_1(\deh_1,\deh_2,\deh_3) \pl_1^\mu+k_2(\deh_1,\deh_2,\deh_3) \pl_2^\mu+k_3(\deh_1,\deh_2,\deh_3) \pl_3^\mu\right] \bar{\psi}(x_1)\hat{\gamma} (iF_{\mu\nu}(x_3)z^\nu) \psi(x_1)\\
&+ k_4(\deh_1,\deh_2,\deh_3) \bar{\psi}(x_1) (iF^{\mu\nu}(x_3)\gamma_{\mu\nu})\hat\gamma \psi(x_2)+k_5(\deh_1,\deh_2,\deh_3) \bar{\psi}(x_1) (iF_{\mu\nu}(x_3)z^\nu) \gamma^\mu \psi(x_2)\,,
\end{aligned}
\end{equation}
where we defined
\begin{equation}
\begin{aligned}
k_1(u,v,w)&=  \frac{2}{w} h(u+w,v)-\frac{1}{w} \big(\frac{3}{2} - (v+w) \pl_u + v \pl_v + w\pl_w\big)g(u,v,w)\,,\\
k_2(u,v,w)&=-\frac{2}{w} \tilde{h} (u,v+w)-\frac{1}{w} \big(\frac{3}{2} + u \pl_u - (u+w) \pl_v + w\pl_w\big)g(u,v,w)\,,\\
k_3(u,v,w)&=\frac{1}{w}( h(u+w,v)
- \tilde{h}(u,v+w)) -\frac{1}{w}  \big(\frac{3}{2} + u \pl_u +v \pl_v -(u+v)\pl_w\big)g(u,v,w)\\
k_4(u,v,w)&= \frac{1}{2}(h(u+w,v) - \tilde{h}(u,v+w))\,,\\
k_5(u,v,w)&=\frac{1}{w}(q(u+w,v)-\tilde{q}(u,v+w))+2\tilde{h}(u,v+w)+\frac{u+v+w}{w} g(u,v,w)\,.
\end{aligned}
\end{equation}
As a check, note that for $s=2$ we are left with $\bar{\psi} F_{\mu\nu}z^\mu\gamma^\nu\psi$, which vanishes upon using the equations of motion.

In 3d Euclidean space, the $\gamma$ matrices are just Pauli matrices, and we have the following identities ($\epsilon_{123}=1$):
\begin{align}
\gamma_\mu \gamma_\nu&= \delta_{\mu\nu}+i\epsilon_{\mu\nu\rho}\gamma^\rho\,,\\
\gamma_{\mu\nu} \gamma_\rho &= i\epsilon_{\mu\nu\rho}+\gamma_\mu \delta_{\nu\rho}-\gamma_\nu \delta_{\mu\rho}\,,\\
\gamma_\mu\gamma_\nu\gamma_\rho&=i\epsilon_{\mu\nu\rho}-\delta_{\mu\rho}\gamma_\nu +\delta_{\nu\rho}\gamma_\mu+\delta_{\mu\nu}\gamma_\rho\,.
\end{align}
Using these, we can write
\begin{align}\notag
\pl \cdot J_{\rm f}&=\left[k_1 \pl_1^\mu+k_2 \pl_2^\mu+k_3 \pl_3^\mu\right] \bar{\psi}(\gamma\cdot z) (iF_{\mu\nu}z^\nu) \psi- k_4 \bar{\psi} F^{\mu\nu}\epsilon_{\mu\nu\rho} z^\rho \psi+(k_5+2k_4) \bar{\psi}(i F_{\mu\nu}z^\nu) \gamma^\mu \psi\,.
\end{align}
Upon using the gauge field equations of motion
\begin{equation}
(F_{\mu\nu})^i_{\ j} =\frac{2\pi}{k}\epsilon_{\mu\nu\rho}\bar\psi_j \gamma^{\rho}\psi^i\,,
\end{equation}
we find
\begin{equation}
\begin{aligned}
\frac{k}{2\pi}\pl \cdot J_{\rm f}&=\left[k_1 \pl_1^\mu+k_2 \pl_2^\mu+k_3 (\pl_3^\mu+\pl_4^{\mu})\right] \bar{\psi}(x_1)\hat{\gamma} i\epsilon_{\mu\nu\lambda}z^\nu (\bar{\psi}(x_3)\gamma^\lambda\psi(x_4)) \psi(x_2)\\
&- 2k_4 \bar{\psi}(x_1)(\bar{\psi}(x_3)\hat\gamma\psi(x_4)) \psi(x_1)+(k_5+2k_4) \bar{\psi}(x_1) i\epsilon_{\mu\nu\rho}z^\nu\gamma^\mu(\bar{\psi}(x_3)\gamma_\rho\psi(x_4)) \psi(x_2)\,.
\label{djf-final}
\end{aligned}
\end{equation}
Note that $\pl_3$ will have to be ``point-split" from now on: $\pl_3 \rightarrow \pl_3 +\pl_4$ (and similarly when $\deh_3$ appears in $k_1\,\ldots, k_5$). To write the result (\ref{djf-final}) as a sum of double-trace primaries, we can use the Fierz identity\footnote{Spinor indices are uncontracted on the left-hand side, so the right-hand side is a 2 by 2 matrix.}
\begin{equation}
\psi\bar\psi = -\frac{1}{2}(\bar\psi \psi)-\frac{1}{2}(\bar\psi\gamma^{\mu}\psi) \gamma_{\mu}\,.
\end{equation}
After using this identity, we can write the descendant as:
\begin{equation}
\begin{aligned}
&\frac{k}{2\pi} \partial\cdot J_{\rm f} = i\epsilon_{\mu\nu \rho}z^{\rho} \Big[k_4 \bar{\psi}_i(x_1) \gamma_{\mu} \psi^i(x_4) \bar{\psi}_j(x_3)
\gamma_{\nu} \psi^j(x_2) \\
&+\frac{1}{2}(k_1 \pl_1^\mu+k_2\pl_2^\mu+k_3(\pl_3^\mu+\pl_4^\mu))
(\bar{\psi}_i(x_1) \hat{\gamma} \psi^i(x_4) \bar{\psi}_j(x_3) \gamma_{\nu} \psi^j(x_2)
+\bar{\psi}_i(x_1) \gamma_{\nu} \psi^i(x_4) \bar{\psi}_j(x_3) \hat{\gamma} \psi^j(x_2) ) \Big]\\
&+(\frac{1}{2}(k_1 \deh_1 +k_2 \deh_2 +k_3 (\deh_3 + \deh_4) + k_5 + 3 k_4) \bar{\psi}_i(x_1) \psi^i(x_4) \bar{\psi}_j(x_3) \hat{\gamma} \psi^j(x_2)\\
&-(\frac{1}{2} (k_1 \deh_1 +k_2 \deh_2 +k_3 (\deh_3 + \deh_4) + k_5 + k_4) \bar{\psi}_i(x_1) \hat{\gamma} \psi^i(x_4) \bar{\psi}_j(x_3) \psi^j(x_2)\,.
\label{djf-Fierz}
\end{aligned}
\end{equation}

It is now convenient to use the following identities, which follow from the free Dirac equation:
\begin{align}
    i\epsilon^{\mu\nu\rho} \pl_{2,4\mu} \gamma_{\nu} = -\partial_{2,4\rho}\,,\\
      i\epsilon^{\mu\nu\rho} \pl_{1,3\mu} \gamma_{\nu} = +\partial_{1,3\rho}\,,
\end{align}
where the subscripts indicate the field we act on, and the sign difference is due to the difference of Dirac equation for $\psi$ and $\bar{\psi}$. We also
have:
\begin{align}
  i\epsilon^{\mu\nu\rho} z_{\rho} \partial_{1,3\mu} \hat{\gamma} = i\epsilon^{\mu\nu\rho} z_{\rho} \gamma_{\mu} \deh_{1,3} + z_{\nu} \deh_{1,3}\,,  \\
  i\epsilon^{\mu\nu\rho} z_{\rho} \partial_{2,4\mu} \hat{\gamma} = i\epsilon^{\mu\nu\rho} z_{\rho} \gamma_{\mu} \deh_{2,4} - z_{\nu} \deh_{2,4}\,.
\end{align}
Using the above identities, we can put (\ref{djf-Fierz}) into the form
\begin{equation}
\begin{aligned}
\partial\cdot J_{\rm f} &=
\frac{2\pi}{k}i\epsilon_{\mu\nu \rho}z^{\rho} \big(k_4 +\frac{1}{2} (k_1 \deh_1 +k_3 \deh_4 -k_2\deh_2 -k_3 \deh_3) \big)\bar{\psi}_i  \gamma_{\mu} \psi^i
\bar{\psi}_j \gamma_{\nu} \psi^j\\
&+\frac{2\pi}{k}( \frac{1}{2}(k_1 \deh_1 +k_2 \deh_2 +k_3 \deh_3 + k_3 \deh_4) + k_5 + 3 k_4 +k_1 \deh_1 -k_3 \deh_4) \bar{\psi}_i \psi^i \bar{\psi}_j \hat{\gamma} \psi^j \\
&+\frac{2\pi}{k}(-\frac{1}{2} (k_1 \deh_1 +k_2 \deh_2 +k_3 \deh_3 + k_3 \deh_4) - k_5 - k_4+k_3 \deh_3 -k_2 \deh_2) \bar{\psi}_i \hat{\gamma} \psi^i \bar{\psi}_j  \psi^j\,.
\label{djf-massage}
\end{aligned}
\end{equation}

To make contact with the decomposition into primaries, it is convenient to define the following object:
\begin{align}
 {\tilde J}^{(s)}_{\mu}= i \epsilon_{\mu\nu}^{\ \ \ \rho} z^\nu j_{\rho}^{(s)}\,,
\end{align}
where on the right-hand side $j_{\rho}^{(s)}$ denotes the spin $s$ current with one free index (and all remaining indices contracted with the
null polarization vector). Using the explicit form of the currents (\ref{jsf}), one can show that
\begin{equation}
{\tilde J}^{(s)}_{\mu} = f_s (\deh_1,\deh_2) \bar{\psi}(x_1) \gamma_{\mu\nu} z^{\nu} \psi(x_2) + \tilde{f}_s(\deh_1,\deh_2) z_{\mu} \bar{\psi}(x_1)\psi(x_2)\,,
\label{mixed}
\end{equation}
where $f_s(u,v)$ is the spin-$s$ part of the generating function in (\ref{jsf}), and $\tilde{f}_s(u,v)$ is given by:
\begin{equation}
\tilde{f}_s(u,v) = \frac{1}{s} \frac{2uv (\pl_u f_s(u,v) - \pl_v f_s(u,v)) +(s-1)(u-v) f_s(u,v)}{u+v}\,.
\end{equation}
The divergence of $\tilde{J}_s$ notably only has the trivial tensor structure:
\begin{equation}
\pl^{\mu} \tilde{J}_{\mu}^{(s)} = ((v-u) f_s + (u+v) \tilde{f}_s) \bar{\psi} \psi\,.
\end{equation}

Then we see that the second and third line of (\ref{djf-massage}) are guaranteed to decompose into products of $\hat{\pl}$-derivatives
of the spin-$s$ currents and
$\hat{\pl}$-derivatives of the scalar operator $\tilde{j}_0=\bar\psi\psi$ or of the divergence $\pl^{\mu} \tilde{J}_{\mu}^{(s)}$.
Noting that $\pl^{\mu} \tilde{J}_{\mu}^{(s)} = i\epsilon^{\mu\nu\rho} z_{\rho} \pl_{\nu} j_{\mu}^{(s)}$,
we see that the second and third line of (\ref{djf-massage}) produce precisely the terms that arise
in the decomposition (\ref{gen-nonconservation}) and (\ref{eq2}). To analyze the terms in
the first line of (\ref{djf-massage}), it is convenient to use explicit light-cone coordinates with $z^{\mu}=\delta^{\mu}_{-}$.
Then one of the $\gamma$-matrices becomes $\gamma_{-}=\hat\gamma$ and the other $\gamma_{3}$. Rewriting $\gamma^3 = i\gamma^{-+}$,
we see from (\ref{mixed}) that the factor $\bar\psi\gamma_{-+}\psi$
has the structure of $\tilde{J}_{-}^{(s)}\sim \epsilon_{-+\rho}j^{\rho}_{(s)}$,
minus the ``scalar-like" term in (\ref{mixed}), that will give rise to terms of the same form as the second and third line of
(\ref{djf-massage}). The end result of the analysis is that (\ref{djf-massage}) precisely takes the form
predicted in section 3:
\begin{equation}
\begin{aligned}
&\partial \cdot J_{\rm f} = \tilde {\cal K}^{(a)}+{\cal K}^{(b)}\,,\\
&\tilde {\cal K}^{(a)} = \sum_{s_1} C_{s_1,\tilde{0},s} [j_{s_1}][\tilde{j}_0]\,,\qquad
{\cal K}^{(b)} = \sum_{s_1,s_2} C_{s_1,s_2,s} [j_{s_1}][j_{s_2}]\,,
\end{aligned}
\end{equation}
where the double-trace operators $[j_{s_1}][\tilde{j}_0]$ and $ [j_{s_1}][j_{s_2}]$ are given respectively in (\ref{eq2}) and (\ref{gen-nonconservation}),
and the overall $C_{s_1, s_2,s}$ coefficients are fixed by our explicit calculation to be
\begin{equation}
C_{s_1,s_2,s} = -\frac{2\pi i\lambda}{N}  \frac{4(s+s_1-s_2)!(s-s_2+s_1)!}{(s+s_1+s_2-1)!(s-s_1-s_2-1)!}\frac{s_1!s_2!}{s!}\,,\quad
s_1+s_2=s-2j\,,\quad j>0
\end{equation}
and
\begin{equation}
C_{s_1,\tilde 0,s} = \frac{2\pi  \lambda}{N} \frac{4(s+s_1) s_1 !}{(s-1)!}\,,\quad s_1=s-2j\,,\quad j>0\,.
\end{equation}
Note that $C_{s_1,s_2,s}$ is the same as for the CS-boson theory (up to the sign), as required by the bose/fermi duality. The result to all
orders in $\lambda$ is obtained using (\ref{C-of-lam}), and was given in (\ref{Csss}) and (\ref{tCss0}).

The form (\ref{djf-final}) (or (\ref{djf-massage})) of the divergence can also be used directly to compute the anomalous dimensions using the master formula
(\ref{mainform}) and the free-fermion propagators. This way we can extract the anomalous dimension to order $\lambda^2$ and exactly in $N$, and we find
\begin{equation}
\gamma_s = \frac{\lambda^2}{N}\left\{0,0,\frac{32}{105}+\frac{8}{105 N},\frac{12}{35}+\frac{4}{105N},
\frac{1504}{3465}+\frac{24}{385 N},\frac{4192}{9009}+\frac{32}{693 N},\ldots\right\}+O(\lambda^3)\,.
\end{equation}
Remarkably, this is identical to the (non-critical) CS-scalar result (\ref{gams-sc}), including the $1/N^2$ term \eqref{subleadB}. Note that setting $N=1$ in these expressions, we obtain the anomalous dimensions in the $U(1)_{k}$ CS-fermion theory to order $1/k^2$.

\subsubsection{Critical fermion}
Let us now study the ``critical" fermionic theory where we add the $(\bar\psi\psi)^2$ interaction in addition to the Chern-Simons gauge field. At least
at large $N$, the theory has a UV fixed point whose $1/N$ expansion can be developed using the action (\ref{crit-fer}). At the UV fixed point,
$\sigma_f$ becomes a scalar primary with $\Delta=1+O(1/N)$, and the $\bar\psi \psi$ operator is formally removed by the $\sigma_f$ equation of motion.

It is evident that the equations of motion for $\bar\psi$ and $\psi$ are modified by terms involving the $\sigma_f$ field.
Omitting terms which are quadratic in the gauge field or $\sigma_f$, the equations of motion are
\begin{equation}
\begin{aligned}
&\slashed{\pl}\psi = i \slashed{A} \psi-\frac{1}{N}\sigma \psi \,,\qquad \pl_{\mu}\bar\psi \gamma^{\mu} = -i\bar\psi \slashed{A}+\frac{1}{N} \sigma\bar{\psi}\,,\\
&\partial^2 \psi=\frac{i}{2} \gamma^{\mu\nu}  F_{\mu\nu}\psi +i (\pl\cdot A)\psi+2i A\cdot \pl \psi-\frac{1}{N}(\pl_{\mu}\sigma)\gamma^{\mu}\psi\,,\\
&\partial^2 \bar{\psi}=\frac{i}{2} \bar{\psi}\gamma^{\mu\nu} F_{\mu\nu} -i\bar{\psi}\pl \cdot A-2 i (\pl^{\mu} \bar{\psi})A_{\mu}
+ \frac{1}{N} \bar{\psi} (\slashed{\pl} \sigma) \,.
\end{aligned}
\end{equation}
The calculation of the divergence then picks up an extra term compared to the ``regular" CS-fermion theory:
\begin{equation}
\begin{aligned}
&\partial \cdot J_{\rm f} = {\cal K}_{\rm reg.CS-fer.}+{\cal K}_{\rm crit.fer.}\,,\\
&{\cal K}_{\rm crit.fer.} =\frac{1}{N}\left[(q(\deh_1+\deh_3,\deh_2)-\tilde{q}(\deh_1,\deh_2+\deh_3)
+ h(\deh_1+\deh_3,\deh_2) - \tilde{h}(\deh_1,\deh_2+\deh_3)\bar{\psi}(x_1)\psi(x_2)\sigma(x_3)\right.\\
&\qquad \qquad
+\left.(h(\deh_1+\deh_3,\deh_2)+\tilde{h}(\deh_1,\deh_2+\deh_3)) \pl^\mu_3 \bar{\psi}(x_1)\gamma_{\mu\nu}z^\nu\psi(x_2) \sigma(x_3)\right]\,,
\end{aligned}
\end{equation}
where $q(u,v)$, $\tilde q(u,v)$, $h(u,v)$ and $\tilde h(u,v)$ were defined in (\ref{nonconsgen}), and ${\cal K}_{\rm reg.CS-fer.}$
is the descendent computed in the previous section, given in (\ref{djf-final}).
After expressing the right-hand side in terms of double-trace primaries, one should impose the condition $\bar\psi\psi=0$. This
amounts to dropping $\tilde{\cal K}^{(a)} \sim \sum_{s_1}C_{s_1,\tilde{0},s} [\tilde{j}_0][j_{s_1}]$ from ${\cal K}_{\rm reg.CS-fer.}$, and
one gets the final result (after also dropping the $\bar\psi \psi$ terms which arise when writing ${\cal K}_{\rm crit.fer.}$ in terms of
primaries):
\begin{equation}
\begin{aligned}
&\partial \cdot J_{\rm f} = {\cal K}^{(a)}+{\cal K}^{(b)}\,,\\
\end{aligned}
\end{equation}
where ${\cal K}^{(b)}\sim \sum_{s_1,s_2}C_{s_1,s_2,s}[j_{s_1}][j_{s_2}]$ remains the same as in the regular CS-fermion theory of the previous section,
and ${\cal K}^{(a)}\sim \sum_{s_1} C_{s_1,0,s}[j_{s_1}][\sigma_f]$ coincides
with the quasi-bosonic result in eq. (\ref{qb-nonconservation}), with $j_0$ replaced by $\sigma_f$, and with the undetermined constants found to be
\begin{equation}
C_{s_1,0,s} = \frac{2i}{N} \frac{(s^2-s_1^2) s_1!}{s!}\,.
\end{equation}
Note that, redefining $\sigma_f = 4\pi\lambda j_0^{\rm crit.fer.}$, this result correctly maps to the divergence in the regular CS-scalar theory,
eq. (\ref{dj-regbos})-(\ref{Ca-regbos}).

\section{Direct Feynman diagram computation}
\label{pert}

In this section, we evaluate the coefficients $a_s$ and $b_s$ by a direct diagrammatic calculation of the anomalous dimensions. $a^F_s$ can be determined by a perturbative calculation in the critical bosonic theory (at $\lambda_b=0$), $a^B_s$ can be determined by a perturbative calculation in the critical fermionic theory (at $\lambda_f=0$). Once $a_s$ is known, then $b_s$ (which must be the same for bosonic and fermionic theories) can be obtained by a two-loop calculation in the non-critical fermionic theory.

In a $U(N_f)_{k_f}$ Chern-Simons theory with fundamental matter, with $k_f$ defined via a dimensional reduction regularization scheme (see \cite{Giombi:2011kc} and \cite{Aharony:2012nh,GurAri:2012is}), $\lambda_f=\frac{N_f}{k_f}$ and $N_f$ are related to $\tilde{\lambda}$, $\tilde{N}$ of\cite{MZ} via:
\begin{eqnarray}
\tilde{N} & = & 2N_b \frac{\sin (\pi \lambda_b)}{ \pi \lambda_b} = 2N_f \frac{\sin (\pi \lambda_f)}{ \pi \lambda_f}	\,, \\
\tilde{\lambda} & = & \tan (\pi \lambda_f/2) = -\cot (\pi \lambda_b/2)\,.
\end{eqnarray}
This implies:
\begin{eqnarray}
\tau^f_s -1 & = & \frac{1}{N_f} \left( \frac{\pi \lambda_f}{4} \right) \tan (\pi \lambda_f/2) \left( a^F_s + b^F_s \cos^2 (\pi \lambda_f/2) \right)\,, \\
& = & \frac{1}{N_f} \frac{\pi^2}{8} \left( a_s^F + b_s^F \right) \lambda_f^2  + O(\lambda_f^4)
\end{eqnarray}
for the two-loop fermionic theory, and
\begin{eqnarray}
	\tau^{cb}_s -1 & = & \frac{1}{2N_b} a_s^F
\end{eqnarray}
for the critical bosonic theory. Identical results hold for the critical fermionic and two-loop bosonic theories.

In section \ref{critical-diagrams}, we include a calculation of $a^B_s$ in the critical fermionic theory, which also appeared earlier in \cite{Muta:1976js}, and in section \ref{two-loop} we include a two-loop calculation of the anomalous dimension in the non-critical fermionic theory to determine $b_s$.

Perturbative calculations of the $1/N$ anomalous dimension for all the higher-spin currents in the critical bosonic theory have been obtained earlier in \cite{ruhl} (see also \cite{Skvortsov:2015pea,Giombi:2016hkj}), so we do not include them here.

We calculate the anomalous dimension of $j_s$ with null polarization vector $z$ for $s \geq 1$. The free vertex for a spin $s$ current with $s \geq 1$, in momentum space can be written as:
\begin{eqnarray}
V_s^0(q,p) & = &  \gamma^\mu z_\mu f_s(i(q-p)\cdot z,ip\cdot z)\,, \\ V_s^0(0,p) & = & \slashed{z} {4^s \over 2s!} (-ip\cdot z)^{s-1} \\
& = & v_s~(p \cdot z)^{s-1} \slashed{z}
\end{eqnarray}
where $f_s$ is determined from the generating function given in equation \eqref{HS-fer}. The anomalous dimension, $\delta_s=\tau_s-1$, of $j_s$ is related to the logarithmic divergence of the corrected vertex $V'(q,p)$ via $V'_s(0,p) =-\delta_s V^0_s(0,p) \log \Lambda$.

\subsection{Critical fermionic theory} \label{critical-diagrams}
We now calculate the $1/N$ anomalous dimension for all the higher-spin currents in the critical fermionic theory. Our conventions are those of \cite{GurAri:2012is}.

The $\sigma$ propagator is:
\begin{equation}
\langle \sigma(q) \sigma(-p) \rangle = G(q)\delta^3(p-q) (2 \pi)^3 = \frac{G_0}{|q|} \delta^3(p-q) (2 \pi)^3 \,,
\end{equation}
where $G_0=8/N$.

\begin{figure}
    \centering
   \includegraphics[width=2cm, height=0.70cm]{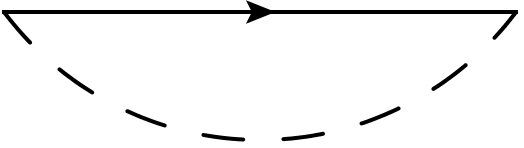}
    \caption{The fermion self-energy correction in the critical fermionic theory.}
    \label{fig:CF1}
\end{figure}

\begin{figure}
    \centering
   \includegraphics[width=1cm, height=1.5cm]{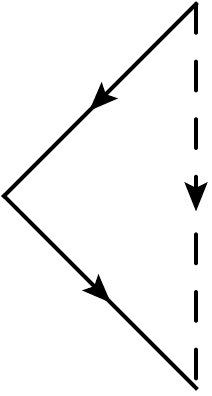}
    \caption{A vertex correction in the critical fermionic theory.}
    \label{fig:CF2}
\end{figure}

\begin{figure}
  \centering
\includegraphics[width=3cm, height=2cm]{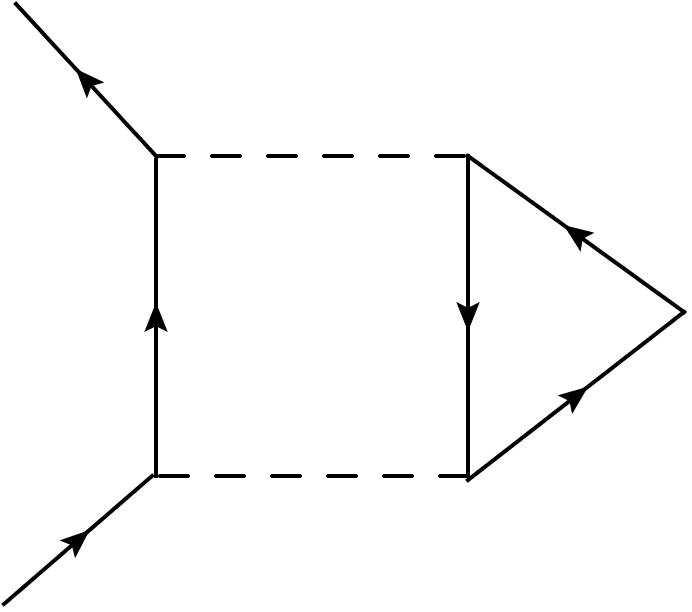}
\includegraphics[width=3cm, height=2cm]{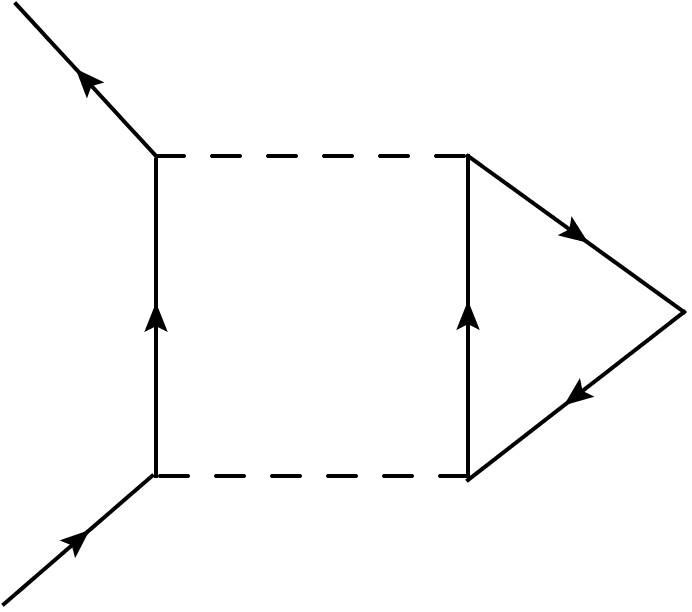}
 \caption{These two diagrams provide a third correction to the vertex in the critical fermionic theory when $s$ is even.}  \label{fig:CF3}
\end{figure}

There are essentially three different diagrams which contribute to the $1/N$ logarithmic divergence of the corrected vertex $V'_s$, depicted in Figures \ref{fig:CF1}, \ref{fig:CF2} and \ref{fig:CF3}.

The fermion self-energy is shown in Figure \ref{fig:CF1}. The logarithmic divergence of the self energy is:
\begin{equation}
\int \frac{1}{i \slashed{p}} G(q-p) \frac{d^3p}{(2\pi)^3}=  - \frac{G_0}{6 \pi^2}~ i\slashed{q} \log \Lambda\,,
\end{equation}
which leads to a contribution of $\frac{G_0}{6 \pi^2}$ to the anomalous dimension.

Another correction to the vertex is shown in Figure \ref{fig:CF2}. The contribution to the corrected vertex $V'_s$ from this diagram is:
\begin{eqnarray}
\hat{V}_s'^{(2)}(0,p) &  = & \int \frac{d^3k}{(2\pi)^3}  G(k) \frac{1}{-i (-\slashed{p}-\slashed{k})} \gamma^\mu z_\mu   f_s(i(-p-k)\cdot z,i(p+k)\cdot z) \frac{1}{i (\slashed{p}+\slashed{k})}  \\
 & = &  n_s \frac{ G_0 }{\pi^2} \log\Lambda  ~\gamma^\mu z_\mu   f_s(-ip\cdot z, ip\cdot z) \nonumber \\
& = & \left( n_s \frac{ G_0 }{\pi^2} \log\Lambda\right)   ~V_s^0(0,p)\,,
\end{eqnarray}
where
\begin{equation}
n_s=\frac{1}{(4s+2)(2s-1)}
\end{equation}

The two diagrams in Figure \ref{fig:CF3} contribute equally to the corrected vertex. Their sum is given by\begin{equation}
V'^{(3)}_s(0,p)=\int \frac{d^3q}{(2\pi)^3} G(q)( A_s(q)+A_s(-q) )G(q) \frac{1}{i \slashed{p}-i\slashed{q}}. \label{an}
\end{equation}
with
\begin{equation}
A_s(q)= -N\text{tr } \int \frac{d^3p}{(2\pi)^3} \frac{1}{i \slashed{p}} V_0(0,p) \frac{1}{i\slashed{p}} \frac{1}{i\slashed{p} - i\slashed{q}}.
\end{equation}

We evaluate
\begin{equation}
A(q)= -v_s \left( \frac{s}{2s-1} \right) i \frac{(2s)!}{4^{s+1} s! s!} \frac{(q\cdot z^s}{q},
\end{equation}
for even $s$ and  $A(q)=0$ for odd $s$.

We find the contribution to the logarithmic divergence from diagram 3 for $s$ even is
\begin{equation}
V_s'^{(3)}(0,p) = 2\frac{G_0^2}{16\pi^2}  \left( \frac{s}{(2s-1)(2s+1)} \right)   V^0_s(0,p) \log \Lambda,
\end{equation}
and $V_s'^{(3)}(0,p) = 0$ for $s$ odd.

Summing all three contributions, the overall logarithmic divergence of the corrected vertex is:
\begin{eqnarray}
\hat{V}'_s & = & \frac{s-2}{6s-3} \frac{8}{N\pi^2} (-\log \Lambda) \hat{V}_s, \text{ for $s$ even, $s>0$.} \\
\hat{V}'_s & = & \left(\frac{2(s^2-1)}{3(4s^2-1)} \right)\frac{8}{N\pi^2} (-\log \Lambda) \hat{V}_s, \text{ for $s$ odd.}
\end{eqnarray}
and the anomalous dimension of the spin $s$ current, with $s>0$ is given by
\begin{equation}
    \tau_s^{\text{critical fermionic}} -1 = \begin{cases} \frac{s-2}{6s-3} \frac{8}{\pi^2} \frac{1}{N}\,, & \text{$s$ even}\,, \\
    \frac{2(s^2-1)}{3(4s^2-1)} \frac{8}{\pi^2} \frac{1}{N}\,, & \text{$s$ odd}\,. \end{cases}
\end{equation}
A similar calculation shows that the anomalous dimension of the scalar primary $\sigma$ is  given by $-\frac{16}{3\pi^2}\frac{1}{N}$, so the above formula does not apply for $s=0$.

\subsection{Two-loop Chern-Simons fermionic theory} \label{two-loop}
Our calculation of two-loop anomalous dimensions closely follows \cite{Giombi:2011kc}.

The higher-spin currents in the interacting, non-critical, fermionic theory are the same as those in the free fermionic theory with all derivatives promoted to covariant derivatives. We calculate anomalous dimensions of $j_{(s)}^{+++\ldots}$ with all upper $+$ indices, in light-cone gauge, $A^+=0$. In this gauge, the generating function for $j_{s}^{+++\ldots}$ is the same as in the free theory, and the vertex contains no factors of $A_\mu$.

In light cone gauge the gauge propagator $\langle A^a_\mu(q) A^b_\nu(-p) \rangle = (2\pi)^3 \delta(q-p) D_{\mu\nu}(q) \delta^{ab}$ is given by:
\begin{equation}
    D_{+3}(q)=-D_{3+}(q)=\frac{4\pi i}{k} \frac{1}{q_-}\,.
\end{equation}

The order $\lambda^2$ correction to the gauge field propagator at $1/N$ is:
\begin{equation}
\begin{pmatrix}
G_{33} & G_{3+} \\
G_{+3} & G_{++}
\end{pmatrix}
= \frac{2\pi^2 \lambda^2}{N^2} q_{+}^2  \frac{1}{q q_s^4}
\begin{pmatrix}
- q_-^2 & q_3 q_- \\ q_3 q_- & q_s^2
\end{pmatrix}
\end{equation}

\begin{figure}
    \centering
    \includegraphics[width=3cm, height=1.5cm]{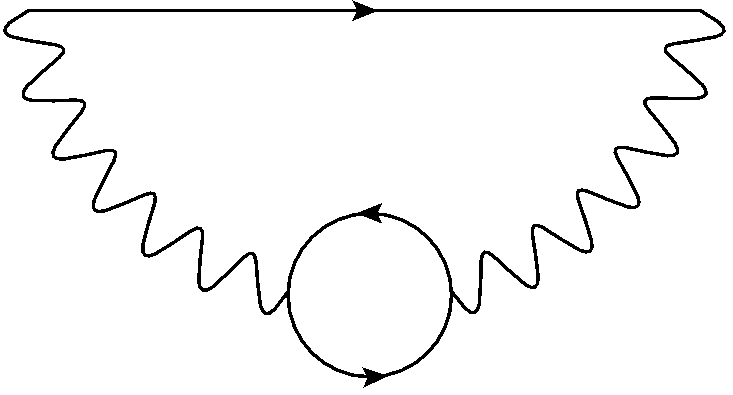}\
    \caption{The two-loop fermion self-energy correction.}
    \label{fig:F1}
\end{figure}

\begin{figure}
    \centering
    \includegraphics[width=2.15cm, height=2cm]{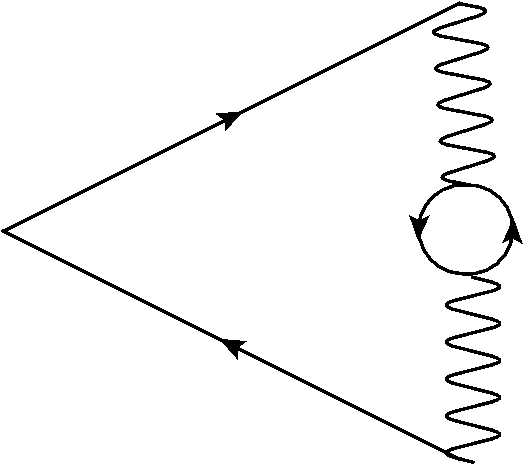}
    \caption{A two-loop vertex correction.}
    \label{fig:F2}
\end{figure}

\begin{figure}
    \centering
    \includegraphics[width=3cm, height=2cm]{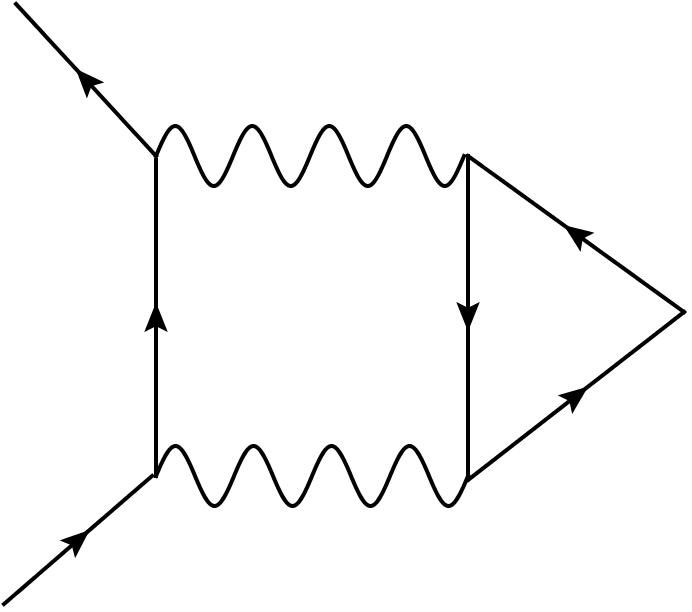}
\includegraphics[width=3cm, height=2cm]{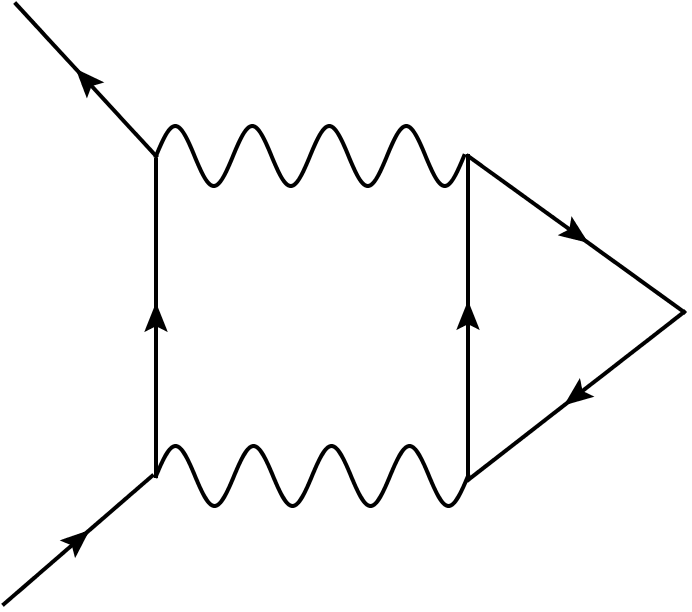}\
    \caption{These two diagrams contribute equally to the two-loop anomalous dimension when $s$ is even.}
    \label{fig:F3}
\end{figure}

Again, there are three diagrams that contribute, depicted in Figures \ref{fig:F1}, \ref{fig:F2} and \ref{fig:F3}.

The contribution of the two loop $1/N$ self-energy of the fermion to the corrected vertex is given by Figure \ref{fig:F1}. Its contribution to the anomalous dimension can be found a two-point function calculation, we find its contribution to the logarithmic divergence of the corrected vertex to be:
\begin{equation}
V'^{(1)}_s = -\frac{11}{24} \frac{\lambda^2}{N}(-\log \Lambda)V^{0}_s.
\end{equation}

The second diagram contributing to the corrected vertex is shown in Figure \ref{fig:F2} and is given by
\begin{equation}
    \frac{N}{2}\int \frac{d^3q}{(2\pi)^3} \left(G_{\mu\nu}(q) \gamma^\mu \frac{1}{i (\slashed{p}+\slashed{q})} v_s \gamma_- (p_-+q_-)^{s-1} \frac{1}{i (\slashed{p}+\slashed{q})} \gamma^\nu \right).
\end{equation}

The contribution to the anomalous dimension can be evaluated via:
\begin{eqnarray}
V'^{(2)}_s & = & V^{0}_s \frac{N}{2}  \frac{1}{2} \Tr \left(\gamma^- \int \frac{d^3q}{(2\pi)^3} \left(G_{\mu\nu}(q) \gamma^\mu \frac{1}{i (\slashed{p}+\slashed{q})} \gamma_- (p_-+q_-)^{s-1}  \frac{1}{i (\slashed{p}+\slashed{q})} \gamma^\nu \right) \right).
\end{eqnarray}
The logarithmic divergence of the integral is
\begin{equation}
    V'^{(2)}_s=  \frac{-1}{4}\frac{\lambda^2}{N} p_-^{s-1} v_s \gamma_- \log \Lambda \left(- \frac{1}{2(4s^2-1)}+g(s) \right)\,,
\end{equation}
where, $g(s)$ is
\begin{equation}
  g(s)=  \gamma -\psi(s)+2\psi(2s) =  \sum_{n=1}^s \frac{1}{n-1/2}.
\end{equation}
Here, $\psi(s)$ is the digamma function. Notice that $g(s) \sim \log s$ for $s$ large.

The last contribution to the corrected vertex is the sum of two diagrams shown in Figure \ref{fig:F3}. Here we evaluate the sum of these diagrams.

The sum of the diagrams is given by:
\begin{equation}
	V'^{(3)}_s =\frac{1}{2}\tr \left( \gamma^- \left(-\frac{N}{2}\right)\int \frac{d^3 q}{(2\pi)^3} \gamma^\mu \frac{1}{i(\slashed{p}-\slashed{q})} \gamma^\nu D_{\mu\alpha}(q) D_{\beta \nu}(q) C^{\alpha \beta}(q)\right)\,,
\end{equation}
where
\begin{equation}
  C^{\mu \nu}(q) = v_s \int \frac{d^3p}{(2\pi)^3} \tr \left( \frac{1}{i (\slashed{p}-\slashed{q})} \gamma^\mu \frac{1}{i \slashed{p}} \gamma_- p_-^{s-1}  \frac{1}{i \slashed{p}} \gamma^\nu \right)
\end{equation}

Evaluating this carefully, we find
\begin{equation}
V'^{(3)}_s(0,p) =  \begin{cases}
 \left(-\frac{2 s^2+1}{4 s^4-5 s^2+1} \right)(\lambda^2/N) (-\log \Lambda) V^0_s(0,p)\,, & \text{$s$ even}\,, \\
 0\,, & \text{$s$ odd}.
\end{cases}
\end{equation}

The anomalous dimension of the spin $s$ current gets contributions from only the first two diagrams for $s$ odd and is:
\begin{eqnarray*}
\tau_{s}-1 & = & \frac{-11}{24}\frac{\lambda^2}{N} +\frac{1}{4}\left(\frac{-1}{2(4s^2 -1)}+g(s)\right)\frac{\lambda^2}{N} \\
 & = & \left(\frac{-11 s^2+2}{6 (4s^2-1)} +\frac{1}{4}g(s) \right)\frac{\lambda^2}{N}\,.
\end{eqnarray*}

The anomalous dimension for even spin currents is:
\begin{eqnarray}
\tau_{s}-1 & = & \frac{-11}{24}\frac{\lambda^2}{N} +\frac{1}{4}\left(\frac{-1}{2(4s^2 -1)}+g(s)\right)\frac{\lambda^2}{N}-\frac{2 s^2+1}{4 s^4-5 s^2+1}\frac{\lambda^2}{N} \nonumber \\
 & = & \left( \frac{-11 s^4+s^2-8}{6 \left(4
   s^4-5 s^2+1\right)}+\frac{1}{4}g(s) \right) \frac{\lambda^2}{N}\,. \label{evenspinresult}
\end{eqnarray}

These anomalous dimensions give rise to the values of $a_s$ and $b_s$ quoted above.

We note that, via a similar calculation, we find that the two-loop anomalous dimension\footnote{We thank Aaron Hui for discussions regarding this calculation.} of the scalar $\tilde{j}_0$ is
\begin{equation}
\tau_{\tilde{0}}-2=-\frac{4}{3}\frac{\lambda^2}{N}.
\end{equation}
which happens to agree with equation \eqref{evenspinresult} when $s \rightarrow 0$.

\section{Constraining the higher-spin symmetry-breaking three-point functions}
\label{three-point-section}

In this section, we use our results for the divergence of $j_s$ from sections \ref{constrain} and \ref{sec:classical} to determine the conformally-invariant, non-conserved parity odd three-point functions $$\langle j_{s_1}(x_1,z_1) j_{s_2}(x_2,z_2) j_{s_3}(x_3,z_3) \rangle.$$ Our analysis in this section uses the results and notation of \cite{Giombi:2011rz}, which we briefly review in appendix \ref{three-point-appendix}, in which conformally invariant three-point functions are expressed in terms of the structures $P_i$, $Q_i$ and $S_i$.

As noted in \cite{Giombi:2011rz}, and in subsequent works \cite{Maldacena:2011jn, MZ}, there exist exactly three conformally invariant conserved structures for $\langle j_{s_1}(x_1,z_1) j_{s_2}(x_2,z_2) j_{s_3}(x_3,z_3) \rangle$ when conservation with respect to all three currents is imposed. These are the free fermion correlation function, the free boson correlation function, and a parity-odd result, unique to three dimensions.

In  \cite{Giombi:2011rz}, based on numerical examples, it was conjectured that the exactly conserved parity odd form exists only when the three spins satisfy the triangle inequality, which takes the form $s_3 \leq s_1+s_2$, if we assume $s_3$ is the largest of the three spins. Below, we prove this result for arbitrary spins.

When the triangle inequality is violated, i.e., $s_3 > s_1+s_2$, a parity-violating form of the three-point function arises in Chern-Simons vector models\cite{MZ} that is conserved with respect to the first two currents only. Requiring the divergence of $j_{s_3}$ to be a conformal primary, we are able to uniquely determine this form; and using the results of the classical divergence calculation, also its correct normalization.

In subsection \ref{scalar-primary-jjj} we present recurrence relations that can be easily solved numerically for the parity odd three-point functions of a scalar operator and two other operators of nonzero spin for correlation functions involving the quasi-fermionic scalar ($\tilde{j}_0$) or the quasi-bosonic scalar $j_0$.

For all spins non-zero, we are able to derive recurrence relations which are valid in a particular limit (the light-like OPE limit of \cite{Maldacena:2011jn}) for arbitrary spins. We are also able to explicitly show that the parity-odd three-point functions are uniquely determined by the divergence of $j_{s_3}$ if the triangle inequality is violated, which implies that, if $j_{s_3}$ is exactly conserved, the parity-odd three-point functions must vanish outside the triangle inequality.

In appendix \ref{tables}, we present some explicit non-conserved parity-odd three-point functions for small spins.

\subsection{Three-point functions involving a scalar primary}

\label{scalar-primary-jjj}
When one of the spins, which we take to be $s_2$, is zero, it is possible to explicitly determine recurrence relations for the three-point functions.

\subsubsection{Quasi-fermionic theory}
The most general ``parity-odd"\footnote{Recall that, by ``parity-odd", here we mean parity different from the free theory, so that the three-point function must be multiplied by an odd power of $\lambda$ when it arises in a Chern-Simons vector model. Three-point functions involving $\tilde{j}_0$ in the theory of free fermions involve an epsilon tensor, and hence the $S_i$'s, and are in this case considered to be ``parity-even".} three-point function involving a parity-odd, twist-two scalar $\tilde{j}_0$ allowed by conformal invariance is:
\begin{equation}
   \langle j_{s_1}(x_1,z_1) \tilde{j}_0 j_{s_3}(x_3,z_3) \rangle = \frac{1}{|x_{12}|^2|x_{23}|^2} \sum_{a=0}^{s_1} \tilde{c}_a Q_1^a (P_2^2)^{s_1-a} Q_3^{s_3-s_1+a}\,, \label{qfsum}
\end{equation}
where the $\tilde{c}_a$ are undetermined coefficients. 

The correlation function is not conserved with respect to $x_3$. Using the results of section \ref{constrain} for $\partial \cdot j_{s} \Big|_{s_1,\tilde{0}}$, we can determine:
\begin{equation}
     \langle j_{s_1}(x_1) \tilde{j}_0(x_2) \partial \cdot {j}_s(x_3) \rangle =  \sum_{m=0}^{p} c_m \partial_-^m \langle j_{s_1}(x_1) j_{s_1}(x_3) \rangle
          \partial_-^{p-m} \langle j_0(x_2) j_0(x_3) \rangle \,,
\end{equation}
which implies,
\begin{equation} \begin{split}
\partial_{(3)}^\mu D^{(3)}_\mu \langle j_{s_1} j_{\tilde{0}} j_{s_3} \rangle & =\frac{(-1)^{s_3-s_1-1}}{ |x_{23}|^4 |x_{31}|^2}  Q_3^{s_3-s_1-1} P_2^{2s_1} \frac{(s_3-s_1)(s_3+s_1-1)!}{2^{2s_1}(2s_1)!} C_{s_1,0,s_3} n_{s_1} n_{\tilde{0}} (1+\tilde{\lambda}^2)\tilde{N}^2\\
& = \frac{\tilde{d}_0}{ |x_{23}|^4 |x_{31}|^2}  Q_3^{s_3-s_1-1} P_2^{2s_1} \label{cons}\,.
\end{split}
\end{equation}

Explicitly evaluating the divergence of Equation \eqref{qfsum} and inserting into equation \eqref{cons} yields a recurrence relation for the $\tilde{c}_a$:
\begin{equation}\begin{split}
   & \tilde{c}_{a-1} \left(4 a^2+a (-6 s_1+2
   s_3-7)+2 s_1
   (s_1+2)-3 s_3+3\right) \\ &
   -
   \tilde{c}_a 2 a(a-s_1+s_3)+\tilde{c}_{a-2}
   (a-s_1-2) (-2 a+2 s_1+3)=0\,,
\end{split}
\end{equation}
which is valid for $2\leq a \leq s_1$, along with the following boundary terms:
\begin{eqnarray}
\tilde{c}_{s_{1}} (s_{1} (2
   s_{3}-1)-s_{3})-\tilde{c}_{s_{1}-1} & = & 0 \\
    (s_{3}-s_{1}) (\tilde{c}_0 (2
   (s_{1}-1) s_{1}-s_{3})+2
   \tilde{c}_1 (s_{1}-s_{3}-1)) & = & \tilde{d}_0\,.
\end{eqnarray}
This recurrence relation ($s_1$ equations in $s_1$ unknowns) has a \textit{unique} solution, which is proportional to $C_{s_1,0,s_3}$. The correlation function therefore necessarily vanishes if $j_{s_3}$ is conserved. It appears to also automatically satisfy conservation with respect to the first current. In Appendix \ref{tables} we present a few solutions to this recurrence relation explicitly.

The reason we are able to solve for the correlation function uniquely is that the number of conformally invariant structures in equation \eqref{qfsum} is independent of the third spin. So, imposing a constraint on the divergence with respect to $s_3$ gives us $s_1$ equations in $s_1$ unknowns, and hence uniquely determines the correlation function.

\subsubsection{Quasi-bosonic theory}

We can write the most general conformally invariant parity-odd correlation function involving a twist-one scalar $j_0$ as:
\begin{equation}
    \langle j_{s_1}(x_1,z_1) j_{0}(x_2) j_{s_3}(x_3,z_3) \rangle= \frac{1}{|x_{12}||x_{23}||x_{31}|} \sum_{a=0}^{s_1-1} \tilde{c}_a Q_1^a (P_2^2)^{s_1-a-1} Q_3^{s_3-s_1+a} S_2 \label{cc}\,,
\end{equation}
where the $\tilde{c}_a$ are undetermined coefficients.

From the constraint that the divergence of $j_{s_3}$ be a conformal primary, we have from section \ref{constrain}:
\begin{equation} \begin{split}
\partial_{(3)}^\mu D^{(3)}_\mu  \langle j_{s_1} j_{0} j_{s_3} \rangle
& = 2^p f_n \frac{(2s_1+n)!(p-n)!}{(2s_1)!} n_{s_1} n_0 (1+\tilde{\lambda}^2)\tilde{N}^2 \left(\epsilon_{\mu\nu-}x_{13}^\mu x_{23}^\nu\right) \\ &
(x_{13}^+)^{2s_1-1+n} (x_{23}^+)^{p-n-1} (x_{23}^2)^{n-p-1} (x_{31}^2)^{-2s_1-n-1}
\\ & =
\frac{|x_{12}|}{|x_{23}|^3 |x_{31}|^3} \tilde{d}_0  Q_3^{s_3-s_1-1} P_2^{2s_1-2} S_2 \,,\label{diveq}
\end{split}
\end{equation}
where $\tilde{d}_0$ is given by
\begin{equation}
\tilde{d}_0= (-1)^{s_3-s_1+1} (1+\tilde{\lambda}^2)\frac{n_{s_1}n_{0}\tilde{N}^2 C_{s_1,0,s_3}(s_3+s_1-1)!}{{2^{2s_1-1}}(2s_1)!} s_1.
\end{equation}

Using equation \eqref{cc}, the equation \eqref{diveq} translates into a recurrence relation for $\tilde{c}_a$.
\begin{equation}
  \begin{split}
  &  \tilde{c}_a  \left(4
     a^2+a (-6 s_{1}+2 s_{3}+3)+2
     (s_{1}-1)
     s_{1}+s_{3}+1\right) \\ & - \tilde{c}_{a-1}
     (2 (a-1)-2 s_{1}+1) (a-s_{1})
     -2 (a+1) \tilde{c}_{a+1}
     (a-s_{1}+s_{3}+1)=0\,,
     \end{split}
\end{equation}
which is valid for $a=1$ to $s_1-2$ along with the boundary terms:
\begin{eqnarray}
   \tilde{c}_0 (s_{3}-s_{1}) (2
     (s_{1}-1) s_{1}+s_{3}+1)-2
     \tilde{c}_1 (s_{3}-s_{1})
     (-s_{1}+s_{3}+1) & = & \tilde{d}_0 \\
    c_{s_{1}-1} (s_{1}
       (2 s_{3}-1)-s_{3}+2)-3
       c_{s_{1}-2} & = & 0\,.
\end{eqnarray}
This has a unique solution ($s_1-1$ equations in $s_1-1$ unknowns), which is proportional to $C_{s_1,0,s_3}$. The correlation function necessarily vanishes if $j_{s_3}$ is conserved. A few solutions to this recurrence relation are given in Appendix \ref{tables}.

\subsection{Three-point functions involving nonzero spins}
Let us briefly consider correlation functions involving all nonzero spins. While it is difficult to say much about these in full generality, following \cite{Maldacena:2011jn}, we work in the ``light-like OPE" limit, which is a constraint on $x_{12}$ and $z_1$ and $z_2$ that commutes with the operation of taking the divergence with respect to $x_3$. In this limit, $P_3=0$ and $S_3=0$; which we shall use frequently in all the derivations below.  One way of taking this limit is to fix the first two polarization tensors to be $z_1^\mu = z_2^\mu = \delta^\mu_-$ and set $x_{12}^+=0$.

In the light-like OPE limit, conformal invariance restricts the parity-odd three-point function to be of the form:
\begin{equation}
\langle j_{s_1}(x_1,z_1) j_{s_2}(x_2,z_2) j_{s_3}(x_3,z_3) \rangle = \frac{1}{|x_{12}| |x_{23}| |x_{31}|} f_{s_1,s_2,s_3}(P_i,Q_i,S_i)
\end{equation}
where
\begin{equation}
\begin{split}
    f_{s_1,s_2,s_3}  = & \sum_{n=0}^{\text{min}(s_3-s_1-1,s_2-1)} \tilde{a}_n Q_2^{s_2-n-1}P_1^{2n} Q_3^{s_3-s_1-n-1} P_2^{2s_1} S_1 \\
    & + \sum_{m=0}^{\text{min}(s_1+s_2-s_3,s_1-1)} \tilde{b}_m Q_1^m Q_2^{s_1+s_2-s_3-m} P_1^{2 (s_3-s_1+m)} P_2^{2(s_1-m-1)}S_2 \\
    & + \sum_{n=\text{max}(0,s_1+s_2-s_3)}^{s_2} \tilde{c}_n Q_2^n Q_3^{s_3-s_1-s_2+n}P_1^{2(s_2-n)} P_2^{2(s_1-1)} S_2 \\
    & + \sum_{n=0}^{\text{min}(s_3-s_2-1,s_1-1)} \tilde{d}_n Q_1^{s_1-n-1} P_2^{2n}Q_3^{s_3-s_2-1-n} P_1^{2s_2} S_2 \label{jjjstructures}.
 \end{split}
\end{equation}
Here we used the constraints (which simplify in the limit $P_3=0$ and $S_3=0$)
\begin{eqnarray}
  Q_1 Q_2 Q_3 & = & Q_1 P_1^2 + Q_2 P_2^2\,, \label{constraint1}\\
  Q_1 S_1 & = & Q_2 S_2\,, \label{constraint2}
\end{eqnarray}
to write the function $f$ in a unique way by eliminating all occurrences of $Q_1Q_2Q_3$ and $Q_1S_1$, so each term is independent, and all powers are positive.

Note that the range of the sums, which was fixed by requiring all exponents to be positive, depends nontrivially on the spins. Let us assume $s_3$ is the largest spin and $s_3\geq s_1+s_2$. Then the total number of undetermined coefficients in \eqref{jjjstructures} is $2s_2+s_1-1$, which is independent of $s_3$.

When we take the divergence with respect to $x_3$ we obtain an expression which is again of the form \eqref{jjjstructures} but with $s_3 \rightarrow s_3'=s_3-1$. However, the number of allowed conformal structures in equation \eqref{jjjstructures} is independent of $s_3$ outside the triangle inequality. This means that, outside the triangle inequality, imposing a constraint on the divergence with respect to $x_3$ gives us $2s_2+s_1-1$ equations in $2s_2+s_1-1$ unknowns. Therefore, we expect that exactly conserved parity-odd three-point functions vanish outside the triangle inequality, at least in the light-like OPE limit. (Inside the triangle inequality, the number of independent terms in equation \eqref{jjjstructures} does depend on $s_3$ so imposing a constraint on conservation with respect to $x_3$ does not uniquely determine the three-point function.)

When $j_{s_3}$ is not conserved, the results for the divergence $\partial \cdot j_{s_3} \Big|_{s_1,s_2}$ in section \ref{constrain} imply the following:
\begin{eqnarray}
    && \langle j_{s_1}(x_1) j_{s_2}(x_2) \partial \cdot j_{s_3}(x_3) \rangle  \label{genthreepoint} \\
    & = & -n_{s_1}n_{s_2}\tilde{N}^2C_{s_1,s_2,s_3}  2^p \frac{(s_3+s_1+s_2-1)!}{p(2s_1)!(2 s_2)!}  \epsilon_{\mu\nu-}x_{13}^{\mu}x_{23}^{\nu}  \nonumber \\ && \times \frac{(x_{23}^{+})^{2s_2-1}}{x_{23}^{4s_2+2}} \frac{(x_{13}^{+})^{2s_1-1}}{x_{13}^{4s_1+2}}
  \left(\frac{x_{23}^{+}}{x_{23}^2}-\frac{x_{13}^{+}}{x_{13}^2}\right)^{p-1} \left(\frac{x_{23}^{+}}{x_{23}^2} A - \frac{x_{13}^{+}}{x_{13}^2} B \right)\,,\nonumber
\end{eqnarray}
where, $p=s_3-s_1-s_2$, 
$ A = s_1(s_1-s_3-s_2)+2 s_1 s_2 (-1)^{s_3+s_1+s_2} $ and
$ B = -2 s_1 s_2 +s_2(s_3+s_1-s_2)(-1)^{s_3+s_1+s_2}$,
which can be written as,
\begin{eqnarray}
    &&   \langle j_{s_1}(x_1) j_{s_2}(x_2) \partial \cdot {j}_{s_3}(x_3) \rangle  = (-1)^{s_3-s_2-s_1}\frac{n_{s_1}n_{s_2}\tilde{N}^2C_{s_1,s_2,s_3}(s_3+s_1+s_2-1)!}{{2^{2s_1+2s_2-1}}(s_3-s_1-s_2)(2s_1)!(2 s_2)!} \frac{|x_{12}|}{|x_{31}|^{3}|x_{23}|^{3}} \nonumber \\
    && \times
    \left(A S_2 P_1^{2s_2}P_2^{2s_1-2} Q_3^{s_3-s_2-s_1-1} +  B S_1 P_2^{2s_1}P_1^{2s_2-2} Q_3^{s_3-s_2-s_1-1}\right),
\end{eqnarray}
in the light-like limit. (If $s_1=s_2$, this equation has to be multiplied by a factor of 2.) This equation translates into a system of linear equations for the various coefficients in equation \eqref{jjjstructures}. This system of equations appears complicated and difficult to solve in general, though one can obtain solutions for particular spins. In Appendix \ref{three-point-appendix}, we change variables to obtain an equivalent, but simpler recurrence relation. Using the new variables, we also prove that exactly-conserved parity-odd three-point functions vanish when $s_3>s_1+s_2$ even outside the light-like limit. We also present a list of some non-conserved parity-odd three point functions with non-zero spins in Appendix \ref{tables}.

\section*{Acknowledgments}

We thank Aaron Hui, Igor Klebanov, Shiraz Minwalla, Eric Perlmutter, Aninda Sinha, Sandip Trivedi, and Ran Yacoby for useful discussions.
The work of SG and VK was supported in part by the US NSF under Grant No.~PHY-1318681 and No.~PHY-1620542.
The work of V.G. and S.P. was supported in part by an INSPIRE award of the Department of Science and Technology, India. The work of E.S. was supported  in part  by the Russian Science Foundation grant 14-42-00047 in association with Lebedev Physical Institute and by the DFG Transregional Collaborative Research Centre TRR 33 and the DFG cluster of excellence ``Origin and Structure of the Universe". E.S., V.K. and S.G. would like to thank Munich Institute for Astro- and Particle Physics (MIAPP) of the DFG cluster of excellence ``Origin and Structure of the Universe" for the hospitality. S.P. would like to thank the Institute for Advanced Study, Princeton; the Centre for High Energy Physics, Indian Institute of Science, Bangalore; and the Tata Institute of Fundamental Research, Mumbai for hospitality.

\appendix
\section{Constraining the divergence of $j_{s}$} \label{appendix-constrain}
To constrain the double-trace operators that can appear on the right hand side of the non-conservation equation by requiring the divergence to be a conformal primary, we use the following commutation relations from the conformal algebra:
\begin{eqnarray}
\left[ M^{\rho \sigma}, P^\mu \right] & = & - i\left( \eta^{\mu\sigma} P^\rho - \eta^{\mu\rho} P^\sigma \right)\,, \\
\left[ K^\nu, P^\mu \right] & = & 2i \left(\eta^{\mu\nu}D+M^{\mu\nu}\right)\,, \\
\left[K^\nu,j_s^{\rho \sigma} \right] & = & 0\,, \\
\left[ M^{\rho \sigma}, j_s^\mu \right] & = & - i\left( \eta^{\mu\sigma} j_s^\rho - \eta^{\mu\rho} j_s^\sigma \right)-i(s-1)\left(\delta^{\sigma}_- j_s^{\mu\rho} - \delta^{\rho}_- j_s^{\mu\sigma} \right)\,, \\
\left[D,j_s\right] & = &-i \Delta_s j_s\,.
\end{eqnarray}
The last three relations express the fact that $j_s$ is a spin $s$ conformal primary with scaling dimension $\Delta_s$. Recall that $\Delta_s=s+1$, except for the quasi-fermionic scalar $\tilde{j}_0$, for which $\Delta_{\tilde{0}}=2$. Here, as in section \ref{constrain}, we are taking all polarization vectors to be given by $z^\mu=\delta^\mu_-$, so $j_s^{\mu\nu} \equiv j_s^{\mu\nu}{}_{---\ldots}$.

A double-trace operator such as $(\partial_-^2 j_{s}) \partial_- j_0$ is proportional to
\begin{equation}
[P_-, ~[P_-, j_s]] ~[P_-,j_0]\,,
\end{equation}
which, using the state-operator correspondence, we can also write schematically as
\begin{equation}
P_-^2 \ket{j_s} P_- \ket{j_0}.
\end{equation}

Let us constrain the double-trace terms in the non-conservation equation involving a quasi-fermionic scalar, given by equation \eqref{eq2} from the main text:
\begin{equation}
   \partial \cdot {j}_s \Big|_{s_1,0} \sim \sum_{m=0}^p c_m P_{-}^m \ket{j_{s_1}} P_-^{p-m} \ket{\tilde{j}_{0}}
   \label{eq2p} .
\end{equation}
where $p=s-s_1-1$.

Acting on this expression with $K_+$, we obtain
\begin{eqnarray}
0 &=& K_+ \partial \cdot {j}_s \Big|_{s_1,0}  \\
   &=&  \sum_{m=0}^p c_m \left( \left( K_+ P_{-}^m \ket{j_{s_1}} \right) P_-^{p-m} \ket{\tilde{j}_{0}} + P_{-}^m \ket{j_{s_1}} \left( K_+ P_-^{p-m} \ket{\tilde{j}_{0}} \right) \right)\\
   & = & \sum_{m=0}^p c_m \left( \left( [K_+ ,P_{-}^m] \ket{j_{s_1}} \right) P_-^{p-m} \ket{\tilde{j}_{0}} + P_{-}^m \ket{j_{s_1}} \left( [K_+,P_-^{p-m}] \ket{\tilde{j}_{0}} \right)\right).
\end{eqnarray}
Then we use
\begin{equation}
    [ K_\delta, P_-^n] = 2 i n P_-^{n-1} \left( \eta_{-\delta} D+M_{-\delta} \right)+2 n (n-1) \eta_{-\delta} P_-^{n-1}
\end{equation}
and the action of the conformal generators on $\ket{j_s}$ to obtain
\begin{equation}
   \sum_{m=1}^p \left(m (m+ 2 s_1) c_m \right) P_{-}^{m-1} \ket{j_{s_1}} P_-^{p-m} \ket{\tilde{j}_{0}} + \sum_{m=0}^{p-1} (m-p) (m-p-1)c_{m} P_{-}^{m} \ket{j_{s_1}} P_-^{p-m-1} \ket{\tilde{j}_{0}}=0,
\end{equation}
which implies
\begin{equation}
 c_m = \frac{-(m-p-1) (m-p-2)}{   m (m+ 2 s_1)}c_{m - 1},
\end{equation}
which can be solved to give equation \eqref{qfdiv}:
\begin{equation}
    c_m  = \left( -1\right)^m \binom{s-s_1}{m} \binom{s+s_1-1}{m+2s_1}  C_{s_1,\tilde{0},s_3}.
\end{equation}
The resulting expression is also annihilated by $K_3$ and $K_-$.

Similar (but more lengthy) calculations determine analogous recurrence relations for contributions to the non-conservation equation involving the quasi-bosonic scalar \eqref{qb-nonconservation} and general non-zero spins \eqref{gen-nonconservation}. These can be solved to give \eqref{qbdiv1}-\eqref{qbdiv2}, and \eqref{gendiv}.

\section{Some results for parity-odd three-point functions}\label{three-point-appendix}
In this appendix, we present slightly simpler recurrence relations for the parity odd three-point functions.

Let us briefly review the notation of \cite{Giombi:2011rz}. Consider the three point function of three operators $\mathcal O_{s_1}(x_1,z_1)$, $\mathcal O_{s_2}(x_2,z_2)$, and $\mathcal O_{s_3}(x_3,z_3)$, of spins $s_1$, $s_2$ and $s_3$ and twists $\tau_1$, $\tau_2$ and $\tau_3$. Conformal invariance restricts the three point function of these operators to take on the form
\begin{equation}
\langle \mathcal O_{s_1} \mathcal O_{s_2} \mathcal O_{s_3} \rangle = \frac{1}{|x_{12}|^{\tau_1+\tau_2-\tau_3}
|x_{23}|^{-\tau_1+\tau_2+\tau_3}
|x_{31}|^{-\tau_2+\tau_1+\tau_3}} f(P_i,Q_i,S_i).
\end{equation}
where $f(P_i,Q_i,S_i)$ is a polynomial in the cross ratios $P_i$, $Q_i$ and $S_i$  defined for $i=1,2,3$, using polarization spinors $z_i^\mu (\sigma_\mu)_{\alpha \beta} \equiv (\lambda_i)_\alpha (\lambda_i)_\beta$ as 
\begin{eqnarray}
P_3 & = & \lambda_1 \frac{\slashed{x}_{12}}{x_{12}^{2}} \lambda_2 \,,\\
Q_3 & = & \lambda_3 \left( \frac{\slashed{x}_{31}}{x_{31}^{2}}+\frac{\slashed{x}_{23}}{x_{23}^{2}}\right) \lambda_3\,, \\
S_3 & = & i \frac{1}{|x_{12}| |x_{23}| |x_{31}|} \left( \lambda_2 \slashed{x}_{12} \slashed{x}_{23} \lambda_3 \right) (\lambda_1 \frac{\slashed{x}_{12}}{x_{12}^{2}} \lambda_2) \,,
\end{eqnarray} 
and cyclic permutations. Here $\slashed{x} \equiv x^\mu \sigma_\mu$. To match spin, $f$ must be homogeneous of degree $s_i$ in each of the $z_i$. The cross ratios are not all independent, and satisfy some constraints listed in \cite{Giombi:2011rz}. 

In terms of the null polarization vectors $z_i$, the cross-ratios can be written as:
\begin{eqnarray}
P_3^2 & = & -2 z_1^\mu z_2^\nu \left( \frac{\delta_{\mu\nu}}{x_{12}^2} - \frac{2x_{12}^\mu x_{12}^\nu}{x_{12}^4} \right)\,, \\
Q_3 & = & 2 z_3^\mu \left( \frac{x_{32}^\mu}{x_{32}^2} - \frac{x_{31}^\mu}{x_{31}^2} \right)\,, \\
S_3 & = & 4 \frac{\epsilon_{\mu\nu\rho}}{|x_{31}||x_{12}|^3 |x_{23}|} \left( x_{12}^\mu x_{31}^\nu z_1^\rho z_2 \cdot x_{12} - \frac{1}{2} ( |x_{31}|^2 x_{12}^\mu + |x_{12}|^2 x_{31}^\mu) z_1^\nu z_2^\rho \right)\,.
\end{eqnarray}
Parity odd three-point functions are linear in the $S_i$, while parity even three-point functions do not contain the $S_i$.

If some of the operators are conserved currents, we must also require that the approriate divergence of the three-point function vanishes.  We note that taking divergences with respect to $x_3$ of a correlation function involving a twist-1 operator is facilitated using the operator $\mathcal{D}_3$ defined in Appendix F of \cite{Maldacena:2011jn}, which satisfies:
\begin{equation}
\begin{split}
& \partial_{\lambda_3} \slashed{\partial}_{x_3} \partial_{\lambda_3}
\frac{1}{|x_{12}|^{\tau_1+\tau_2-1}
|x_{23}|^{-\tau_1+\tau_2+1}
|x_{31}|^{-\tau_2+\tau_1+1}}
f(P_1,P_2,P_3,Q_1,Q_2,Q_3)
\equiv \\ &
 \frac{1}{|x_{12}|^{\tau_1+\tau_2-3}
 |x_{23}|^{-\tau_1+\tau_2+3}
 |x_{31}|^{-\tau_2+\tau_1+3}} \mathcal{D}_3 f(P_1,P_2,P_3,Q_1,Q_2,Q_3).
 \end{split}
\end{equation}
In the main text, we defined the divergence of $j_s$ using  $\partial \cdot j_s(x,z) \equiv  \partial_{x^\mu} D^\mu_{z} j_s(x,z)$. A useful relation is
\begin{equation}
\partial_\lambda \slashed{\partial}_x \partial_\lambda = 4 \partial^\mu D_\mu.
\end{equation}

\subsection{A simpler form for the recurrence relations}
\label{solution}

Equation \eqref{jjjstructures} for $\langle j_{s_1} j_{s_2} j_{s_3} \rangle$ can also be written as:
\begin{equation}
\langle j_{s_1} j_{s_2} j_{s_3}(x_3,z_3) \rangle = \frac{1}{|x_{12}| |x_{23}| |x_{31}|} \sum_{a=0}^{s_3-1} \left( c_a Q_1^{s_1-1-a}P_2^{2a}P_1^{2(s_3-1-a)} Q_2^{s_2-s_3+1+a} S_2 \right). \label{general}
\end{equation}
after using the identities: $Q_3=P_1^2/Q_2 + P_2^2/Q_1$ and $Q_1 S_1=Q_2 S_2$ to eliminate $Q_3$ and $S_2$.  To fix the range of $a$ we note that, starting from a polynomial including $S_2$ and $Q_3$ with all non-negative exponents, after using identities to eliminate $Q_3$ and $S_1$, we could end up with an expression where the exponents of the $Q_i$ are negative; however the exponents of the $P_i$ must still be non-negative. (Note that $c_a$ defined here is unrelated to the $c_a$ that appears in section \ref{constrain} or Appendix \ref{appendix-constrain}.)

While any three-point function of the form \eqref{jjjstructures} can be written in the form \eqref{general}, not every expression in the form \eqref{general} corresponds to a valid three-point function. To see this, note that equation \eqref{general} can also be rewritten as
\begin{equation}
\langle j_{s_1} j_{s_2} j_{s_3}(x_3,z_3) \rangle = \frac{1}{|x_{12}| |x_{23}| |x_{31}|} \sum_{m=0}^{s_3-1} \tilde{c}_m Q_1^{s_1-1-m} Q_3^{s_3-1-m} P_2^{2m} Q_2^{s_2}S_2\,, \label{cor1}
\end{equation}
where
\begin{equation}
\tilde{c}_m = \sum_{a=0}^m (-1)^{s_3-1-n} \binom{s_3-1-a}{s_3-1-m}c_a\,. \label{c-tilde}
\end{equation}
If $s_2=0$, then even outside the light like OPE limit, the correlation function must be of the form \eqref{cor1}. If $s_3>s_1$, which we assume in what follows, we must have $\tilde{c}_m=0$ for $m>s_1-1$. This is an extra constraint on the $c_n$.

For all spins nonzero,  there are also constraints on $\tilde{c}_m$ that arise from demanding that the expression can be written in terms of only positive powers of the various cross-ratios $P_1$ , $P_2$, $Q_1$, $Q_2$ and $Q_3$, $S_1$ and $S_2$. To obtain one such constraint, which is sufficient for our purposes, choose points and polarization spinors so that $Q_3=0$ which implies $Q_1 P_1^2=-Q_2P_2^2$. Then, outside the triangle inequality, equation \eqref{jjjstructures} vanishes. However, \eqref{general} does not vanish unless
\begin{equation}
\sum_{a=0}^{s_3-1} (-1)^a c_a =0 \,,\label{extraconstraint}
\end{equation}
which must be imposed for \eqref{general} to represent a valid three-point function, outside the triangle inequality. (Inside the triangle inequality, we do not need to impose \eqref{extraconstraint}.)

\subsubsection*{Conserved three-point functions}

To take the divergence with respect to $x_3$, we act on the above expression with the operator $\mathcal D_3$ derived in equations F.2 of \cite{Maldacena:2011jn}.

In the limit $P_3=0$, for expressions independent of $Q_3$, it takes the simple form (equation I.4 of \cite{Maldacena:2011jn}):
\begin{equation}
\begin{split}
\mathcal{D}_3 = & - (1+2P_1 \partial_{P_1} + 2Q_2 \partial_{Q_2})Q_1 \partial_{P_2}^2 + (1+2P_2 \partial_{P_2} + 2Q_1 \partial_{Q_1})Q_2 \partial_{P_1}^2 \\
& + \left(P_3^2 \partial_{Q_2} +2 P_2P_3 \partial_{P_1} \right) \partial_{P_2}^2 -\left( P_3^2 \partial_{Q_1} + 2 P_1P_3 \partial_{P_2}\right) \partial_{P_1}^2\,.
\end{split}
\end{equation}
We also use the identities 2.20 of \cite{Giombi:2011rz} to derive $S_1^2=Q_2 P_1^2 P_2^2 Q_1^{-1}$, which yields:
\begin{equation}
\partial_{P_1} S_1 = S_1 P_1^{-1}, ~ \partial_{P_2} S_1= S_1 P_2^{-1}, ~ \partial_{Q_1} S_1 = -\frac{1}{2} S_1 Q_1^{-1}, ~ \partial_{Q_2} S_1 = \frac{1}{2} S_1 Q_2^{-2}.
\end{equation}
These relations are valid only when $P_3=S_3=0$, i.e., in the light-like limit.

We find
\begin{equation}
\begin{split}
& \mathcal D_3 \sum_{a=0}^{s_3-1} \left( c_a Q_1^{s_1-a}P_2^{2a}P_1^{2(s_3-1-a)} Q_2^{s_2-s_3+a} S_1 \right) = \\
& \sum_{a=0}^{s_3-2} 4 \left(- c_{a+1} (s_3+s_2-a-1)(a+1)(2a+3)+c_a (1+a+s_1)(2s_3-2a-1)(s_3-a-1) \right) \chi
\end{split}
\end{equation}
where
\begin{equation}
\chi=Q_1^{s_1-a} P_2^{2a}P_1^{2s_3-2a-2}Q_2^{s_2-s_3+1+a}S_1.
\end{equation}

If $j_{s_3}$ is exactly conserved, then the condition that the divergence of equation \eqref{general} is equal to $0$ gives rise to the following recurrence relation:
\begin{equation}
\frac{c_{a+1}}{c_a} = \frac{ (1+a+s_1) (2s_3-2a-1) (s_3-a-1)}{(s_3+s_2-a-1) (a+1) (2a+3)}. \label{rec}
\end{equation}
This has a unique solution for any values of $s_1$, $s_2$ and $s_3$. It can be expressed in terms of Pochhammer symbols as
\begin{equation}
  c_a = -\frac{(-1)^a c_0 (s_{1}+1)
   (s_{3}-1) (2 s_{3}-1)
   (s_{1}+2)_{a-1}
   \left(\frac{3}{2}-s_{3}\right)_{a-
   1} (2-s_{3})_{a-1}}{3 (2)_{a-1}
   \left(\frac{5}{2}\right)_{a-1}
   (s_{2}+s_{3}-1)
   (-s_{2}-s_{3}+2)_{a-1}}
 \end{equation}
and the sum is a hypergeometric function
\begin{equation}
  \langle j_{s_1} j_{s_2} j_{s_3}(x_3,z_3) \rangle = \frac{Q_1^{s_1-1}P_1^{2(s_3-1)} Q_2^{s_2-s_3+1} S_2}{|x_{12}| |x_{23}| |x_{31}|}    c_0  ~
     {}_3F_2\left(s_1+1,\frac{1}{2}-s_3,1-s_3;\frac{3}{2},-
     s_2-s_3+1;u\right)
\end{equation}
where $u=-\frac{P_2^2Q_2}{Q_1P_1^2}$.

If $s_3>s_1+s_2$, then, as discussed above, we must also impose the extra constraint \eqref{extraconstraint} for our solution to represent a valid three-point function expressible in the form \eqref{jjjstructures}. This is
\begin{equation}
c_0 \,
   _3F_2\left(s_{1}+1,\frac{1}{2}-s_3,1-s_{3};\frac{3}{2},-s_2-s_{3}+1;1\right) =0
\end{equation}
which implies that $c_0=0$ and the exactly conserved correlation functions vanish outside the triangle inequality, in the light-like OPE limit. (In section \ref{general_triangle_inequality}, we argue that these correlation function vanish even outside the light-like OPE limit.)

\subsubsection*{Non-conserved parity odd three-point functions}

Outside the triangle inequality, the parity-odd three-point function is not conserved.  In the light-like limit, its divergence with respect to $x_3$ takes the form:
\begin{equation}
 \partial_{x_3^\mu} D^\mu_{z_3} \langle j_{s_1}(x_1) j_{s_2}(x_2) j_{s_3}(x_3,z_3) \rangle =
\frac{|x_{12}|}{|x_{23}|^3 |x_{31}|^3} \sum_a d_a Q_1^{s_1-a-1}P_2^{2a} P_1^{2s_3-2a-4}Q_2^{s_2-s_3+2+a}S_2 \label{divergence}
\end{equation}
where,
\begin{equation}
{d_a} = - c_{a+1} (s_3+s_2-a-1)(a+1)(2a+3)+c_a (1+a+s_1)(2s_3-2a-1)(s_3-a-1) \label{a}
\end{equation}

The result of the divergence calculation, \eqref{genthreepoint} determines the $d_a$ in terms of $C_{s_1,s_2,s_3}$:
\begin{eqnarray}
d_{s_1-1} & = & A K\,, \\
d_{s_1-1+n} & = & \left( \binom{p-1}{n}A + \binom{p-1}{n-1}B \right) K\,, \\
d_{s_1-1+p} & = & d_{s_3-s_2-1}  = B K\,,
\end{eqnarray}
with all other $d_a=0$, and we define $$K = (-1)^{s_3-s_2-s_1}\frac{n_{s_1}n_{s_2}\tilde{N}^2C_{s_1,s_2,s_3}(s_3+s_1+s_2-1)!}{{2^{2s_1+2s_2-1}}(s_3-s_1-s_2)(2s_1)!(2 s_2)!}\,.$$ The spin-dependent constants $A$ and $B$ were defined below equation \eqref{genthreepoint}.

We can now determine a recurrence relation the $c_a$ in terms of $d_a$ (and hence $C_{s_1,s_2,s_3}$) using equation \eqref{a}. Equation \eqref{a} can be written as:
\begin{equation}
c_{a+1} = E c_a + F d_a\,, \label{recurrence}
\end{equation}
where
\begin{equation}
E= \frac{(1+a+s_1)(2s_3-2a-1)(s_3-a-1)}{(1+a)(2a+3)(s_3+s_2-a-1)}, ~ F=-\left((1+a)(2a+3)(s_3+s_2-a-1) \right)^{-1}.
\end{equation}

The solution to equation \eqref{a} for $c_a$ depends on two parameters: $c_0$ and $C_{s_1,s_2,s_3}$ (which enters through the $d_a$), but imposing the extra constraint \eqref{extraconstraint} determines the $c_0$ in terms of $C_{s_1,s_2,s_3}$. Alternatively, we can obtain a relation between $c_0$ and $C_{s_1,s_2,s_3}$ by demanding conservation with respect to the other currents, before taking the light-like limit.

\subsubsection*{Quasi-fermionic scalars}

For parity-odd correlation functions involving the (parity-odd) twist-two quasi-fermionic scalar operator, we have:
\begin{equation}
    \langle j_{s_1} j_{\tilde{0}} j_{s_3} \rangle = \frac{1}{|x_{12}|^2 |x_{23}|^2} \sum_{a}^{s_3-1} c_a Q_1^{s_1-a} P_2^{2a} P_1^{2(s_3-a)} Q_2^{-s_3+a}.
\end{equation}
and, using
\begin{equation}
     \tilde{\mathcal D}_3 = -(2+2P_1 \partial_{P_1} + 2 Q_2 \partial_{Q_2} ) Q_1 \partial_{P_2}^2 + (2P_2 \partial_{P_2} + 2 Q_1 \partial_{Q_1} )Q_2 \partial^2_{P_1}.
\end{equation}
we can write its divergence as,
\begin{equation}
    \partial_{x_3^\mu} D^\mu_{z_3}
    \langle j_{s_1} j_{\tilde{0}} j_{s_3}(x_3,z_3) \rangle = \frac{1}{ |x_{23}|^4 |x_{31}|^2} \sum_{a=0}^{s_3-1} d_a Q_1^{s_1-a} P_2^{2a} P_1^{2(s_3-a-1)} Q_2^{-s_3+a+1}.
\end{equation}
with
\begin{equation}
    d_a=(s_3-a)(s_1+a)(2s_3-2a-1)c_a-(a+1)(2a+1)(s_3-a)c_{a+1}.
\end{equation}

Comparing to our earlier expression \eqref{cons}, we find
\begin{equation}
    d_{s_1+n} = \binom{s_3-s_1-1}{n} (-1)^{s_3-s_1-1} \frac{(s_3-s_1)(s_3+s_1-1)!}{2^{2s_1}(2s_1)!} n_{s_1}n_{\tilde{0}} C_{s_1,\tilde{0}}
\end{equation}
for $n=0,~\ldots,s_3-s_1-1$.

The recurrence relation depends on two unknown parameters: $c_0$ and $C_{s_1,0,s_3}$. By requiring the correlation function to vanish when $Q_3=0$, we obtain a relation between these two parameters. 


\subsection{Conserved parity-odd three-point functions vanish outside the triangle inequality} \label{general_triangle_inequality}

Above, we showed that the conserved parity-odd three-point function vanishes outside the triangle inequality in the light-like limit, where $P_3=0$ and $S_3=0$. Let us extend this to a proof that the conserved parity-odd three-point functions vanishes even outside the light-like limit.

Our strategy is to expand the three-point function as a power series in $P_3$, where we count $S_3 \sim P_3$. Let $m>0$, and use induction. If we assume all terms of order $P_3^{m-1}$ vanish, we can show that conservation implies the terms of order $P_3^m$ must also vanish.

The most general three-point function can be written as:
\begin{equation}
\langle j_{s_1} j_{s_2} j_{s_3}(x_3,z_3) \rangle = \frac{1}{|x_{12}| |x_{23}| |x_{31}|} f(P_i,Q_i,S_i).
\end{equation}
By writing down the most general allowed form for $f(P_i,Q_i,S_i)$ that is order $P_3^{2m}$, and dropping terms of order $P_3^{2m+1}$ and higher we can see that, if $s_3>s_1+s_2$, any term we write down must be proportional to $Q_3^z$, where $z\geq 1$. Hence the correlation function must vanish in the limit that $Q_3=0$. 
 
We can also write the three-point function in terms of $Q_1$, $Q_2$, $P_1$, $P_2$ only, by explicitly solving the constraints in \cite{Giombi:2011rz} and allowing negative exponents for the $Q_i$:
\begin{equation}
\langle j_{s_1} j_{s_2} j_{s_3}(x_3,z_3) \rangle = P_3^{2m} \frac{1}{|x_{12}| |x_{23}| |x_{31}|} \sum_{a=0}^{s_3-1} \left( c_a Q_1^{s_1-m -1-a}P_2^{2a}P_1^{2(s_3-1-a)} Q_2^{s_2-m-s_3+1+a} S_2 \right). \label{general2}
\end{equation}
Then, imposing conservation, we find an essentially identical recurrence relation to Equation \eqref{rec} above. When we also impose the condition that it vanishes when $Q_3=0$, as in equation \eqref{extraconstraint}, we find that there is no solution.

\section{List of parity-odd three-point functions}

\label{tables}

We present the non-zero parity-odd three-point functions outside the triangle inequality, for spins up to 6 in the theory, using values of $C_{s_1,s_2,s_3}$ derived from the classical equations of motion.

\paragraph{Correlation Functions involving a Quasi-Fermionic Scalar}
The correlation functions listed below are to be multiplied by
$$\tilde{N} {\tilde{\lambda}} \frac{1}{|x_{12}|^2 |x_{23}|^2}.$$ (We remind the reader that our normalization for the scalar, given in equation \eqref{normalization-j0}, is such that this is exact to all orders in $\tilde{\lambda}$.)

\begin{eqnarray*}
\langle j_1 j_{\tilde{0}} j_3 \rangle & = &
\frac{Q_3^2 \left(2 P_2^2+Q_1
   Q_3\right)}{8 \pi ^4}
   \\
 \langle j_2 j_{\tilde{0}} j_4 \rangle & = &
-\frac{Q_3^2 \left(-10 P_2^2
   Q_1 Q_3+P_2^4-Q_1^2
   Q_3^2\right)}{8 \pi ^4}
    \\
 \langle j_1 j_{\tilde{0}} j_5 \rangle & = &
\frac{Q_3^4 \left(4 P_2^2+Q_1
   Q_3\right)}{8 \pi ^4}
    \\
 \langle j_3 j_{\tilde{0}} j_5 \rangle & = &
-\frac{ Q_3^2 \left( P_2^4
   Q_1 Q_3-22 P_2^2 Q_1^2
   Q_3^2+16 P_2^6-Q_1^3
   Q_3^3\right)}{8 \pi ^4}
    \\
 \langle j_2 j_{\tilde{0}} j_6 \rangle & = &
\frac{ Q_3^4 \left(16 P_2^2
   Q_1 Q_3+4 P_2^4+Q_1^2
   Q_3^2\right)}{8 \pi ^4}
     \\
 \langle j_4 j_{\tilde{0}} j_6 \rangle & = &
-\frac{ Q_3^2 \left(-24
   P_2^6 Q_1 Q_3-125 P_2^4
   Q_1^2 Q_3^2-38 P_2^2 Q_1^3
   Q_3^3+62 P_2^8-Q_1^4
   Q_3^4\right)}{8 \pi ^4}
\end{eqnarray*}

\paragraph{Correlation functions involving a Quasi-Bosonic Scalar}
The correlation functions listed below are to be multiplied by $$\tilde{N}{\tilde{\lambda}} \frac{1}{|x_{12}||x_{23}||x_{31}|}.$$
\begin{eqnarray}
\langle j_1 j_0 j_3 \rangle & = & -\frac{i Q_3^2 S_2}{16 \pi ^4}
\\
\langle j_2 j_0 j_4 \rangle & = & -\frac{i Q_3^2 S_2 \left(4
   P_2^2+Q_1 Q_3\right)}{16
   \pi ^4}
\\
\langle j_1 j_0 j_5 \rangle & = & -\frac{i Q_3^4 S_2}{16 \pi ^4}
\\
\langle j_3 j_0 j_5 \rangle & = & -\frac{i Q_3^2 S_2 \left(2
   P_2^2+Q_1 Q_3\right)
   \left(6 P_2^2+Q_1
   Q_3\right)}{16 \pi ^4}
\\
\langle j_2 j_0 j_6 \rangle & = & -\frac{i Q_3^4 S_2 \left(6
   P_2^2+Q_1 Q_3\right)}{16
   \pi ^4}
\\
\langle j_4 j_0 j_6 \rangle & = & -\frac{i Q_3^2 S_2 \left(107
   P_2^4 Q_1 Q_3+40 P_2^2
   Q_1^2 Q_3^2+102 P_2^6+3
   Q_1^3 Q_3^3\right)}{48 \pi
   ^4}
\end{eqnarray}

\paragraph{All Spins nonzero}
The correlation functions listed below are to be multiplied by $$\tilde{N} \frac{\tilde{\lambda}}{1+\tilde\lambda^2} \frac{1}{|x_{12}||x_{23}||x_{31}|}.$$
We also omit an overall numerical normalization constant (which is in principle determinable from $C_{s_1,s_2,s_3}$ and the recurrence relations given above.) These are all valid outside the light-like OPE limit as well. To fix the coefficients of terms that vanish in the light-like limit, we also imposed conservation with respect to $x_1$ and $x_2$.
\begin{eqnarray}
\langle j_1 j_2 j_5 \rangle & \sim &
 Q_3^2 \Big(-6  P_1^4  S_2+6  P_1^2  Q_1
    Q_3  S_1+15  P_1^2  Q_2  Q_3  S_2- P_1^2  Q_3^2
    S_3+5  P_2^2  Q_2  Q_3  S_1+ P_3^2  Q_3^2
    S_1+ \nonumber \\ && Q_2  Q_3^3  S_3 \Big)  \\
    \langle j_1 j_1 j_6 \rangle & \sim &
   {  Q_3^5 (3 Q_1  S_1+3 Q_2  S_2-2 Q_3
   S_3)} \\
\langle j_1 j_3 j_6 \rangle & \sim &
Q_3^2 \Big(72  P_1^6  S_2-32  P_1^4  Q_1
    Q_3  S_1-208  P_1^4  Q_2  Q_3  S_2+9  P_1^4  Q_3^2
    S_3-148  P_1^2  P_2^2  Q_2  Q_3  S_1 \nonumber \\ && -58  P_1^2  Q_2
    Q_3^3  S_3+26  P_1  P_2  P_3  Q_2  Q_3^2  S_1-18
    P_2^2  Q_2^2  Q_3^2  S_1-6  P_3^2  Q_2  Q_3^3  S_1  \nonumber \\ 
    && - 3
    Q_2^2  Q_3^4  S_3 \Big)  \\
\langle j_2 j_2 j_6 \rangle & \sim &
 Q_3^3 \Big(14  P_1^4  Q_1  S_2-88  P_1^3
    P_2  P_3  S_2+95  P_1^2  P_2^2  Q_2  S_2+27  P_1^2
    P_2^2  Q_3  S_3+7  P_1^2  Q_1^2  Q_3  S_1 \nonumber \\ && +7  P_1
    P_2  P_3  Q_2  Q_3  S_2-39  P_1  P_2  P_3  Q_3^2
    S_3+21  P_2^4  Q_2  S_1+7  P_2^2  Q_2  Q_3^2  S_3 \nonumber \\ && +3
    P_3^2  Q_3^3  S_3 \Big)
\end{eqnarray}

\bibliographystyle{ssg}
\bibliography{CSBib}

\end{document}